\documentclass[aps,prb,amsmath,amssymb,superscriptaddress,reprint, a4paper,nobibnotes] {revtex4-1}
\usepackage{graphicx}
\usepackage{hyperref,amsmath}
\usepackage{braket,amsmath}
\usepackage{dcolumn}
\usepackage{bm}
\usepackage{newfloat}
\usepackage{color}

\begin{document}

\title{Free spectral range electrical tuning of a high quality on-chip microcavity}

\author{Christiaan Bekker}
\thanks{CGB and CB contributed equally \\ Corresponding author: c.baker3@uq.edu.au}
\affiliation{Centre for Engineered Quantum Systems, School of Mathematics and Physics, The University of Queensland, Australia}
\author{Christopher G. Baker}
\thanks{CGB and CB contributed equally \\ Corresponding author: c.baker3@uq.edu.au}
\affiliation{Centre for Engineered Quantum Systems, School of Mathematics and Physics, The University of Queensland, Australia}
\author{Rachpon Kalra}
\affiliation{Centre for Engineered Quantum Systems, School of Mathematics and Physics, The University of Queensland, Australia}
\author{Han-Hao Cheng}
\affiliation{Centre for Engineered Quantum Systems, School of Mathematics and Physics, The University of Queensland, Australia}
\affiliation{Centre for Microscopy and Microanalysis, The University of Queensland, Australia}
\author{Bei-Bei Li}
\affiliation{Centre for Engineered Quantum Systems, School of Mathematics and Physics, The University of Queensland, Australia}
\author{Varun Prakash}
\affiliation{Centre for Engineered Quantum Systems, School of Mathematics and Physics, The University of Queensland, Australia}
\author{Warwick P. Bowen}
\affiliation{Centre for Engineered Quantum Systems, School of Mathematics and Physics, The University of Queensland, Australia}


\begin{abstract}
Reconfigurable photonic circuits have applications ranging from next-generation computer architectures to quantum networks, coherent radar and optical metamaterials. However, complete reconfigurability is only currently practical on millimetre-scale device footprints.
Here, we overcome this barrier by developing an on-chip high quality microcavity with resonances that can be electrically tuned across a full free spectral range (FSR). FSR tuning allows resonance with any source or emitter, or between any number of networked microcavities. 
We achieve it by integrating nanoelectronic actuation with strong optomechanical interactions that create a highly strain-dependent effective refractive index.
 This allows low voltages and sub-nanowatt power consumption.
 We demonstrate a basic reconfigurable photonic network, bringing the microcavity into resonance with an arbitrary mode of a microtoroidal optical cavity across a telecommunications fibre link. Our results have applications beyond photonic circuits, including widely tuneable integrated lasers, reconfigurable optical filters for telecommunications and astronomy, and on-chip sensor networks.
\end{abstract}




\maketitle


Dynamically reconfigurable photonic circuits are expected to have a rich variety of applications. For instance, enabling high-bandwidth optical interconnects and memories in next generation computer architectures~\cite{atabaki2018integrating, kuramochi2014large}, chip-based quantum networks~\cite{konoike2016demand, elshaari2017chip, aoki_observation_2006}, and on-chip coherent radar and microwave communication systems~\cite{ghelfi2014fully, xue2018microcomb, zhuang2007single}. Widely tuneable high quality microcavities are a key component for such circuits. Their passive response allows controllable optical phase shifts~\cite{zhuang2007single}, memories~\cite{kuramochi2014large} and add-drop filters~\cite{klein_reconfigurable_2005} which together provide the reconfigurability of the circuit; while their strong optical confinement enhances light-matter interactions and thereby enables components such as lasers~\cite{polman2004ultralow,li2013low}, sensors~\cite{lu2011high, li2014single, forstner2014ultrasensitive, heylman_optical_2017}, optical frequency combs~\cite{del2007optical}, and quantum processors~\cite{aoki_observation_2006}.  

Full reconfigurability requires that the optical resonance frequencies of each microcavity are tuneable over at least half a free spectral range (FSR), since this allows the interaction of any two spectrally narrow components regardless of their initial frequencies. It is then possible to envisage not only fully-reconfigurable photonic circuits, but also arrays of microcavities forming dynamically controlled optical metamaterials~\cite{shalaev2007optical} or on-chip microsensor networks~\cite{iqbal2010label}, and to study collective phenomena such as phase-transitions and topological behaviour in networks of strongly interacting nonlinear photonic systems~\cite{heinrich2011collective, douglas2015quantum, gil-santos_light-mediated_2017}. Furthermore, FSR tuneable microcavities have many other possible applications. 
For example, they could allow widely tuneable on-chip lasers~\cite{polman2004ultralow,li2013low}, resonant coupling of arbitrary microcavity modes to low-linewidth solid-state and fibre laser sources, reconfigurable filters for background rejection and spectroscopic measurements in astronomy~\cite{ellis_photonic_2017}, matching of cavity resonance frequencies to narrow atomic resonances in cavity quantum electrodynamics~\cite{aoki_observation_2006}, and stabilisation of on-chip optical frequency combs~\cite{del2007optical}. 
Despite this range of applications, it has proved challenging to achieve full FSR tuning in a manner that is scalable and allows sub-millimetre device footprints. Approaches demonstrated to-date would require either raising the microcavity to a prohibitively high temperature~\cite{armani_electrical_2004} or straining it more than is possible with standard piezoelectric materials~\cite{jin_piezoelectrically_2018, nguyen_large_2017}.
 
\begin{figure*}[ht!]
 \centering
 \includegraphics[width=\linewidth]{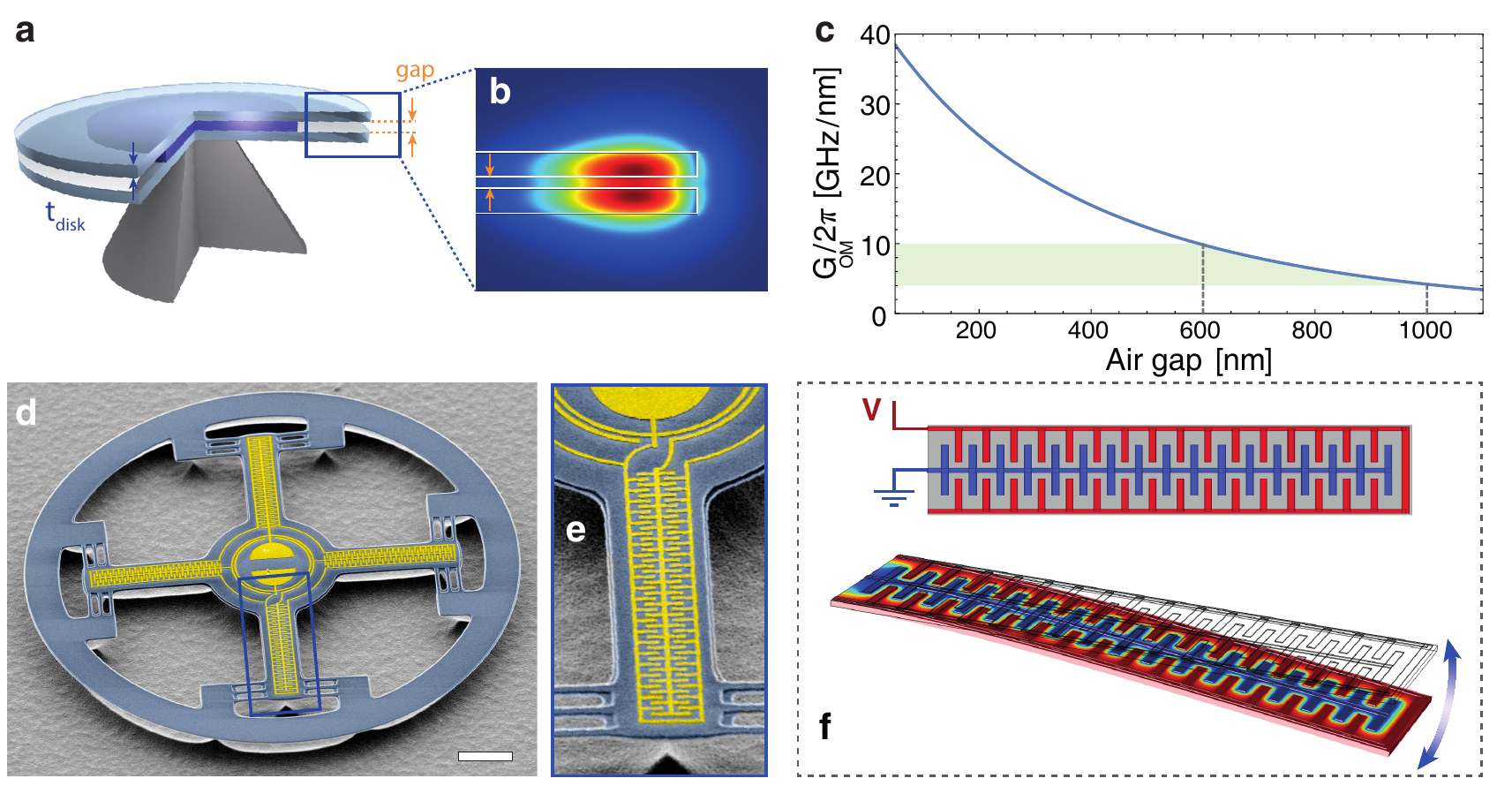}
 \caption{\label{Figure1}  (a) Schematic illustration of a double-disk resonator, consisting of two disks of thickness $t_{\mathrm{disk}}$ (blue) separated by a thin sacrificial layer (purple), which, once etched, reveals an air gap where the optical mode is localized. (b) Finite Element Method (FEM) simulation showing a cross-section  of a fundamental transverse electric (TE) Whispering Gallery Mode (WGM) of the structure (orange arrows highlight the position of the air gap). (c) FEM calculation of the optomechanical coupling strength $G_\text{OM}\equiv\frac{\partial \omega}{\partial x}$ for a silica double-disk with $t_{\mathrm{disk}}=350$ nm, and the WGM shown in (b), as a function of air gap. Note that unlike single-disk resonators \cite{ding_high_2010}, $G_\text{OM}$ is essentially independent of device radius and only depends on the vertical separation between the disks \cite{lin_mechanical_2009}. The shaded area denotes the range of air gaps typically observed in fabricated devices.
 (d) False-color Scanning Electron Microscope (SEM) top-view of a fabricated silica (blue) double-disk device (radius 90 $\mu$m), supported by four spokes. Half-circular gold (yellow) pads in the center of the device are contact pads for the probe tips \cite{baker_high_2016, bekker_injection_2017}. Scale bar is 20 $\mu$m. (e) False-color SEM micrograph showing a zoomed-in view of the gold interdigitated electrodes patterned on the support spokes of the top disk. (f) Top: top-view of a simulated support spoke. The center electrode (blue) is kept at ground, while a nonzero potential bias is applied to the outer electrode (red).  Bottom: 3D FEM electromechanical simulation showing the deflection of the cantilever spoke through the capacitive drive. Color code shows the electric potential (blue=0V; red=1V). Note that for a single-disk resonator only the in-plane change in the  length of the spoke would be useful, as only it changes the cavity radius \cite{baker_high_2016, jin_piezoelectrically_2018}, while the double-disk geometry allows the much larger out-of-plane motion to be leveraged for tuning.} 
 \end{figure*} 
 
 Here we address this challenge, reporting electrical FSR tuning of a  high quality silicon chip-based optical microcavity. 
The key advance is to combine strong nanoelectronic actuation with an engineered microcavity structure that exhibits a highly strain-dependent effective refractive index. The effective refractive index is engineered using the strong optomechanical interaction provided by a double-disk microcavity~\cite{rosenberg_static_2009, wiederhecker2009controlling}. Hybridisation of the  modes of the disks creates an effective index that is strongly dependent on the disk separation, which itself is controlled via electrostatic actuation provided by integrated interdigitated capacitors. 
Using this approach, we demonstrate the ability to tune optical resonances by up to 9~GHz/V$^2$. This allows FSR tuning with applied voltages of less than 15~V, and a full tuning range in excess of three FSRs.
Our devices are fabricated from silica-on-silicon, offering a wide transparency window across and beyond the telecommunications band, but could easily be transferred to other materials such as silicon nitride, silicon-on-insulator, or complementary metal-oxide-semiconductor (CMOS)~\cite{atabaki2018integrating}. 
Capacitive actuation facilitates ultralow power operation, compatible with scalable photonic circuits --- the microcavity can be held resonant at any frequency within the silica transparency window using less than a nanowatt of electrical power. To demonstrate  the broad-tuning capabilities, we implement a simple two microcavity reconfigurable photonic network, showing that the double-disk microcavity can be brought into resonance with an arbitrary fixed-frequency microtoroidal optical cavity across a telecommunications fibre link. 

\subsection*{Background}

Most techniques which enable broad tuning of optical cavities can be sorted into two categories. The first applies heat to the cavity through a laser \cite{zhang_synchronization_2012}, metal probe \citep{armani_electrical_2004} or integrated microheater \cite{klein_reconfigurable_2005,lee_-chip_2017}. This causes a change in temperature $\Delta T$ of the cavity which modifies its effective refractive index through the thermo-optic effect. The second utilises strain caused by an applied force to deform the boundary of the cavity~\cite{sumetsky_super_2010,pollinger_ultrahigh-_2009}. To date, full FSR tuning has not proved possible on sub-millimeter footprints using either of these approaches, although millimeter-scale on-chip silicon-nitride ring resonators have recently been reported with FSR strain tuning provided by an integrated piezoelectric element~\cite{jin_piezoelectrically_2018}. In an alternative approach, FSR tuning has been reported with a split-ring microcavity \cite{chu_wide_2014}, consisting of two evanescently coupled curved waveguides. In this case, physically splitting the cavity allows increased mechanical compliance and therefore improved tunability, but introduces inherently large losses that strongly limit the optical quality factor. 

In general, the resonance condition of an optical cavity of round-trip physical length $L$ requires the optical path length to be an integer multiple $m$ of the free space wavelength $\lambda_0 $:
\begin{equation}
n_{\mathrm{eff}} \,L =m \,\lambda_0
\label{Eq1}
\end{equation}
where $n_{\text{eff}}$ is the effective refractive index of the cavity. FSR tuning to an adjacent longitudinal mode of the cavity ($m \rightarrow m\pm 1$) requires the optical path length to be modified by the free space wavelength. Heat based tuning accomplishes this through a change in the effective refractive index $\Delta n_\text{eff}$, while strain based tuning changes the physical size $\Delta L=\epsilon L$ of the cavity, where $\epsilon$ is the resultant strain from the applied force. From Eq. \ref{Eq1}, these tuning mechanisms result directly in the conditions for FSR tuning: 
\begin{equation}
\epsilon_\text{FSR} = \frac{\lambda}{n_\text{eff}\,L}   \qquad
\Delta n_\text{eff, FSR} = \frac{\lambda}{L} \; \; . \label{Eq2}
\end{equation}
The inverse cavity length scaling present in both cases explains why it is highly challenging to achieve FSR tuning for microscale cavities. For instance, tuning a silica disk with radius $R=100 \; \mu$m by an FSR would require $\Delta T_\text{FSR}$ upwards of 200$^\circ$C, or a radial strain of $\epsilon_\text{FSR} \sim 0.2 \; \%$, exceeding the maximum strain of common piezoelectric materials such as PZT (lead zirconium titanate) \cite{nguyen_large_2017}  (see supplementary information for more information). Moreover, heat-based tuning suffers from two additional drawbacks, namely typically slow thermal response times and power consumption typically upwards of several milliwatts per device to achieve and maintain the large temperature increases required for significant tuning \citep{lee_-chip_2017}. Alternatively, refractive-index tuning could be achieved by electro-optic techniques with materials such as lithium niobate. However, while these enable ultrafast modulation rates up to tens of Gb/s \cite{chen_hybrid_2014, wang_nanophotonic_2018}, they typically allow for much smaller tuning ranges of only a few optical linewidths, so are not included further in this discussion.

In this work, we overcome this miniaturization bottleneck with a technique which brings FSR-tuning capabilities to high quality microscale devices. Whereas most thermal-based approaches rely on changing the effective refractive index  $n_{\mathrm{eff}}$ through the material's thermo-optic coefficient, and most strain-based approaches rely on changing the cavity length $L$ (see Eq. \ref{Eq1}), it is possible to use the optomechanical interaction to engineer an effective refractive index $n_{\mathrm{eff}}$ which is very strongly strain-dependent, much beyond the intrinsic photoelastic properties of the material \cite{baker_photoelastic_2014,balram2014moving}. This allows far greater tunability to be observed than that achievable through simple physical compression of the cavity.

\subsection*{Double-disk geometry}

Previous approaches to refractive index engineering have typically involved bringing an external dielectric into the near-field of the cavity \cite{errando-herranz_low-power_2015}. Here we achieve greatly enhanced refractive index shifts by engineering the strain to modulate the coupling between two optical resonances. As an additional advantage, in this configuration any scattered light is preferentially scattered back into the optical supermode formed by the coupled cavities. This minimises energy loss when compared with the introduction of an external dielectric. We employ a double-disk Whispering Gallery Mode (WGM) geometry, previously reported by several groups \citep{wiederhecker2009controlling,jiang_high-q_2009, lin_mechanical_2009, rosenberg_static_2009, wiederhecker_broadband_2011, zhang_synchronization_2012}. Such a cavity consists of two several-hundred-nanometer thick stacked disks, separated by a thin sacrificial layer which is etched-out to leave an air gap (Fig. \ref{Figure1}(a)). The optical field is shared between both disks in a supermode, with a significant part of the energy situated in the gap (Fig. \ref{Figure1}(b)). This makes the resonance wavelength very sensitive to changes in the separation between the disks, which corresponds to a large optomechanical coupling strength $G_\text{OM}=\frac{\partial \omega}{\partial x}$ \cite{aspelmeyer_cavity_2014}, see Fig. \ref{Figure1}(c).  
This large coupling strength, combined with the much greater compliance of the disk resonators to out-of-plane deflection rather than purely radial compressive strain\footnote{To estimate this, we can compare the typical mechanical resonance frequency $\Omega_M$ of the radial breathing mode (tens of MHz) for devices of this size to that of the out-of-plane flexural mode (tens of kHz). With the spring constant scaling as $k \propto \Omega_M^2$, this gives roughly six orders of magnitude larger compliance.}, allows for the use of much smaller forces to tune the device. For example, even the optical gradient force (radiation pressure) due to circulating power in the resonator can be used to achieve significant tuning of the optical resonance frequency \cite{rosenberg_static_2009, thourhout_optomechanical_2010}, including full FSR tunability \cite{wiederhecker_broadband_2011}. With this purely optical tuning technique, however, a several milliwatt widely tunable pump laser is required for every microcavity to achieve and maintain the desired resonance frequency. The need for multiple tunable lasers, and associated power consumption, precludes use in a scalable photonic network or circuit. Here, instead, we apply tuning forces by integrating interdigitated capacitive electrodes with sub-micron characteristic dimensions onto the surface of the cavity. This provides a direct, scalable and low power electronic tuning mechanism. 

\section{Results}

\subsection*{Fabrication and device design}

Figure \ref{Figure1}(d) shows a SEM top-view of a fabricated double-disk electro-optomechanical cavity.
Devices are fabricated from a wafer containing two silica (SiO$_2$) layers with nominal thickness $t_{\mathrm{disk}}=350$ nm separated by an amorphous silicon ($\alpha$-Si) sacrificial layer of nominal thickness 300 nm, grown by ICP-CVD atop a silicon substrate, see Fig. \ref{Figure1}(a). We use reactive ion etching to etch through the three-layer stack (SiO$_2$/$\alpha$-Si/SiO$_2$) with an electron-beam lithography (EBL)-defined pattern in the shape of an annulus supported by four wide spokes. We release the device through removal of the $\alpha$-Si sacrificial layer with a XeF$_2$ isotropic dry etch. The released annuli confine light in WGM resonances and have outer diameter of 180 $\mu$m; more than a factor of six smaller than the smallest previous high quality FSR-tunable on-chip cavity \cite{jin_piezoelectrically_2018}. Thin tethers are included at the point of attachment of the spokes with the annulus to minimize buckling effects due to residual compressive stress in the silica layers coming from the deposition process. Gold sub-micron  interdigitated electrodes (500 nm width and spacing) are patterned on the top disk, covering the surface of the support spokes (see Fig. \ref{Figure1}(e)). The top and bottom disks are free to move independently, effectively modulating the air gap distance and the structure's effective refractive index in response to a voltage bias on the drive electrodes. We choose this design as a compromise that provides large available area to deposit electrodes for capacitive driving and minimizes buckling effects (see supplementary information for more details and full process flow). We note that the design can be modified to include integrated optical waveguides \cite{baker_critical_2011} and electrical bridges to the contact pads \cite{lee_-chip_2017}, to enable fully on-chip packaging.  We also note that the capacitive tuning method presented here is quite general and material agnostic. It would work with any combination of optically transparent and sacrificial layer materials such as GaAs/AlGaAs, Si/SiO$_2$, SiN/SiO$_2$ or SiN/Si, and is CMOS compatible.

\begin{figure} [t!]
\centering
\includegraphics[width=.95\linewidth]{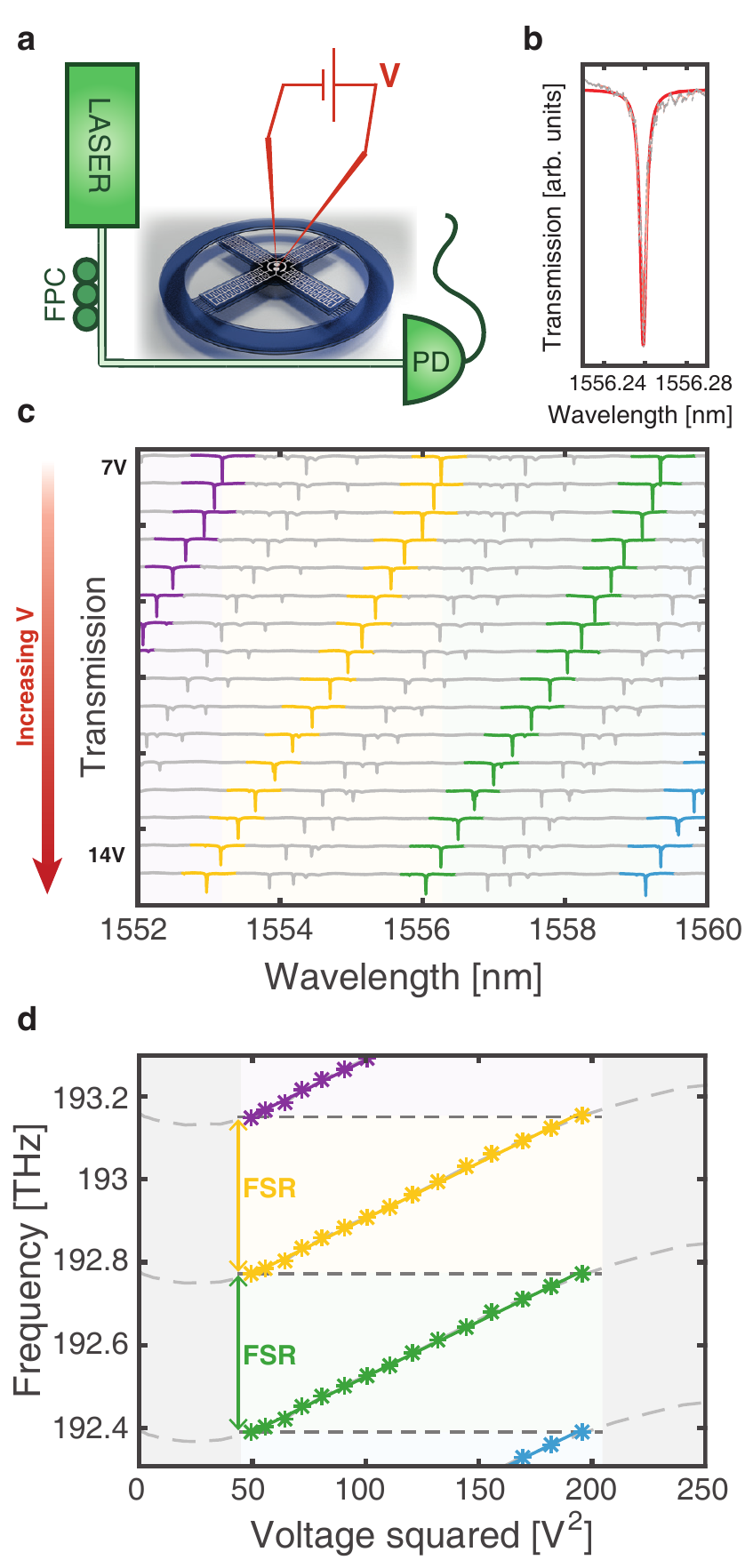}
\caption{\label{Fig2}(a) Experimental setup. Probe tips contacting the device are shown in red. FPC=fibre polarization controller; PD = Photodetector. (b) Typical transmission spectrum of a WGM resonance, measured by fast single-pass sweep of the laser. Quality factor of the fitted Lorentzian curve (red) is $3.8\times 10^5$. For averaged measurements, thermal motion of the disks induces wavelength jitter on the order of 10 pm \cite{rosenberg_static_2009}. (c) Waterfall plot showing full FSR capacitive tuning of a double-disk resonator. Consecutive traces are offset by 0.5 V, with an initial applied voltage of 7 V. Four WGMs of the same family and increasing azimuthal order $m$ are highlighted in purple, yellow, green and blue. (d) WGM resonance frequencies as a function of $V^2$, for the four WGMs highlighted in (c). Slope of the curves corresponds to $\alpha_\text{opt}$.}
 \end{figure}

\subsection*{Modelling of capacitive tuning} 

For capacitive tuning, and assuming linear mechanical response, the optical frequency shift of the cavity is given by $\Delta \mathrm{f}=\alpha_\text{opt}\,V^2$, where
the optical tunability $\alpha_\text{opt}$ can be expressed as:
\begin{equation}
\alpha_\text{opt} = \frac{G_\text{OM}}{2\pi} \alpha_\text{mech}.
\label{Eqalphaopt}
\end{equation}
Here, the mechanical tunability $\alpha_\text{mech}= \frac{1}{k}\times\frac{1}{2} \left(\frac{\partial C}{\partial x}\right)$ is a measure of the physical compliance of the structure times the efficiency of the capacitive actuation along the vertical direction, with $k$ the spring constant of the structure. The mechanical tunability quantifies the physical change in disk separation in response to an applied voltage, with the total deflection in the vertical direction given by $x=F_\text{cap}/k=\alpha_\text{mech}\,V^2$, where the applied capacitive force $F_\text{cap}=\frac{1}{2} \left(\frac{\partial C}{\partial x}\right)$.

In order to estimate the mechanical tunability of the double-disk cavity, we make the simplifying assumption that any vertical motion of the support spokes will be transferred directly to the outer annulus, thereby changing the separation of the disks. In this approximation, the mechanical tunability of the entire device is equal to that of the individual spokes. We therefore seek to quantify how the spokes react to a voltage applied on the interdigitated electrodes. 
Since the response to an applied voltage involves not only uniaxial compression of the spoke, but also significant out-of-plane deflection and change in the spoke's curvature, the analytical estimation of $\alpha_\text{mech}$ is non-trivial. For this reason, we perform 3D Finite Element Method (FEM) electromechanical simulations (see Fig. \ref{Figure1}(f)) to calculate the new equilibrium position of the spoke with an applied voltage bias. These predict a mechanical tunability of the support spokes on the order of $\alpha_\text{mech}\simeq$ 1 nm/V$^2$. 

To quantify the optomechanical coupling strength for a given disk thickness we need only to know the size of the air gap between the disks. Due to variation in material stress and conditions during fabrication, the gap size was experimentally found to vary over several hundred nanometers. In Figure \ref{Figure1}(c) the typical range of disk separations observed after fabrication is shown, with a corresponding range of coupling strengths of $G_\text{OM}/2\pi \in [4;10]$ GHz/nm. Combined with the predicted mechanical tunability, Eq. \ref{Eqalphaopt} yields a range of predicted optical tunabilities of $\alpha_\text{opt}\in [4;10]$ GHz/V$^2$ (see supplementary material for more information). For a double-disk structure of 90 $\mu$m outer radius, this predicts FSR tuning to be easily achievable with applied voltages of less than 15~V.

These predictions indicate that refractive index engineering along with interdigitated capacitors can overcome the scaling bottleneck for resonator tuning, allowing electrical FSR tuning of microscale devices. It is notable, however, that they are far from the ultimate limit of this technique. With further optimisation of both the double-disk geometry and capacitive actuation, the device radius could be reduced to below 15 $\mu$m while maintaining FSR-tuning capability with an applied voltage of less than 15~V (see supplementary information). We note that piezoelectric actuation, such as utilised in the work of Jin, \textit{et al.}\cite{jin_piezoelectrically_2018}, could be used as an alternative to interdigitated capacitors. Based on our predictions, capacitive actuation, beyond the advantage of requiring only a single deposition step to fabricate, interestingly also provides more efficient tuning for devices at the scale of tens of microns (see supplementary information).

\subsection*{Free spectral range tuning}

Figure \ref{Fig2}(a) shows a schematic of the experimental setup. Light is evanescently coupled to the double-disk through a tapered optical fibre. Spectroscopy of the devices is performed with a tunable diode laser (Yenista T100S-HP). The two interdigitated electrodes are contacted using ultrasharp tungsten probe tips, which are connected to a DC voltage source. (See supplementary information for more details.) Figure \ref{Fig2}(b) shows a typical optical resonance of a device, measured by sweeping the laser frequency rapidly over the resonance, yielding an intrinsic optical quality factor of $3.8\times 10^5$. 

Figure \ref{Fig2}(c) displays a series of successive optical transmission spectra of a device as the WGMs of the device are tuned across an entire FSR. Consecutive traces from top to bottom are acquired as the voltage $V$ is ramped up by steps of 0.5 V, starting from $V=7$ V, with full FSR tuning achieved at $V = 14$ V. We observe that the optical quality factor of the modes remain unperturbed throughout the entire tuning range. Next, the frequencies of the four highlighted WGMs in Fig. \ref{Fig2}(c) are tracked versus $V^2$, in order to extract the optical tunability $\alpha_\text{opt}$. These results are shown in Fig. \ref{Fig2}(d). In the region $V^2\in$ [50; 200],  corresponding to the voltage range $V\in$ [7; 14] shown in Fig. \ref{Fig2}(c), we observe a linear trend, with a slope of $\alpha_\text{opt}=3$ GHz/V$^2$. Comparable tuning is found in other devices of a similar design, with a maximum observed tunability of 9 GHz/V$^2$ and tuning ranges exceeding three free spectral ranges (see supplementary information). These results are in line with the predictions from the simulations outlined above, where uncertainty in the disk separation led to a prediction of $\alpha_\text{opt}\in [4;10]$ GHz/V$^2$ . In addition to the expected $V^2$ scaling, further evidence that the tuning is capacitive in nature (and not for instance due to electrostatic interactions between the top disk electrodes and trapped charges on the bottom disk) is provided by observing that the tuning has the same direction with a positive e.g. (0; +10 V) as with a negative e.g. (0; -10 V) bias applied to the electrodes.

Outside of the range $V\in$ [7; 14], the relationship between frequency shift and the square of the applied voltage departs from linearity, implying a changing optical tunability. We ascribe this behaviour to nonlinearities in the mechanical response due to out-of-plane warping of the double-disk due to residual stress in the silica layers, as confirmed by optical profilometry measurements\footnote{We note here interestingly that SEM measurements cannot be relied on for accurate measurements of the disks' geometry and separation, because the significant charging brought about by the electron beam creates strong electrostatic forces that modify the disk separation and can cause collapse of the double-disk structure.} of the fabricated double-disks (see supplementary materials). This warping causes mechanical buckling transitions which modify the mechanical compliance, sometimes with clear steps in the optical tunability (see supplements). 

As well as allowing full FSR tunability, a second key feature of the capacitive tuning presented in this paper is its ultra-low power consumption. In contrast to thermal tuning techniques which require the heating to be maintained as long as the wavelength offset is required, here the only power required to maintain a wavelength offset is that dissipated through leakage current in the electrodes. We measure a leakage current of $\sim$2 pA per Volt applied to the electrodes, corresponding to a leakage/parasitic resistance of $\sim$500 G$\Omega$. With an applied voltage of $V=14$ V required to maintain full FSR tuning, this parasitic resistance results in a power consumption of $\sim$400 pW. This ultralow power consumption makes our approach particularly scalable.

\begin{figure} [b!]
\centering
\includegraphics[width=\linewidth]{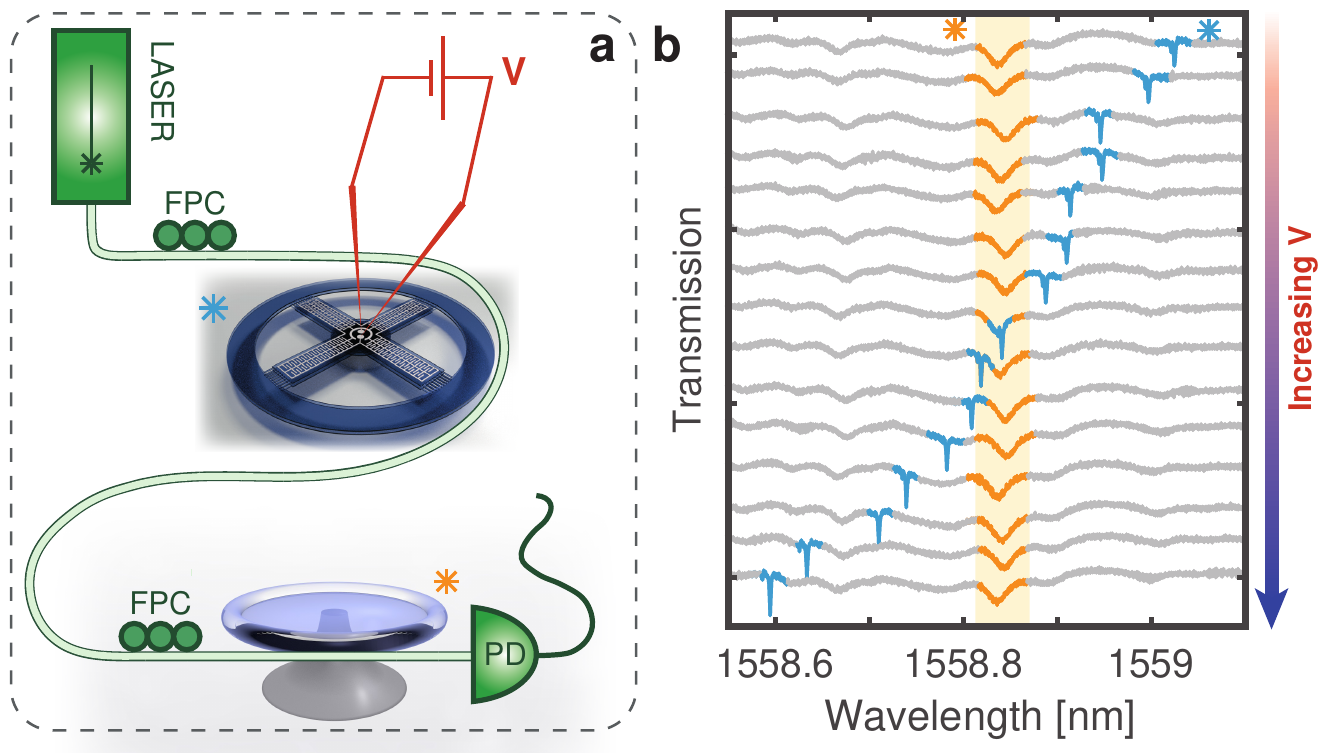}
\caption{\label{Fig3} (a) Schematic of the setup used for coupling between a tunable double-disk resonator and a passive microtoroid cavity. (b) The lower quality toroid WGM (yellow) remains stationary while the optical resonance  of the double-disk (blue) is tuned into resonance with --and then through-- the toroid WGM.}
\end{figure}

\subsection*{Demonstration of a basic reconfigurable photonic network}

As a demonstration of the capability of full FSR microcavity tuning, we employ it in a simple reconfigurable  photonic network. In this network, a tunable double-disk cavity is coupled to an arbitrary optical mode of a passive microtoroidal cavity through a telecommunications optical fibre link,  as shown in Fig. \ref{Fig3}(a).
Initially, the optical resonances of the two devices are far separated in frequency, as seen in the top trace of Fig. \ref{Fig3}(b). Indeed, two resonators taken at random have very low chances of sharing a common resonance, particularly as devices dimensions are scaled down and for high optical quality factors, with the odds scaling as the reciprocal of the product of their finesses.  The resonance of the microtoroid remains static, determined by its geometry, while applying a voltage to the double-disk allows it to be tuned into resonance with the toroid. This allows for switching between interacting and non-interacting cavities at will \cite{zhang_synchronization_2012, lee_-chip_2017}, but also mimics the coupling of arrays of dissimilar cavities for optical networks or optomechanical synchronization \cite{gil-santos_light-mediated_2017, zhang_synchronization_2012}, the tuning of a cavity to an atomic transition in cavity QED \cite{aoki_observation_2006}, or the coupling of a resonator to a fixed-wavelength laser source.

\section{Discussion}

We have reported full FSR electrical tuning of a high quality silicon chip-based optical microcavity. To achieve this we develop a new approach to FSR tuning, combining engineering of the optomechanical interaction to create a highly strain-dependent effective refractive index with electrical actuation through integrated interdigitated capacitors. This approach overcomes a key bottleneck for  nanophotonic circuit-compatible FSR tuning, avoiding the need to locally raise the microcavity temperature by hundreds of degrees~\cite{armani_electrical_2004} or apply strains larger than are available from standard piezoelectric materials~\cite{jin_piezoelectrically_2018, nguyen_large_2017}. We demonstrate frequency tuning over more than three FSRs, with tuning over an FSR requiring applied voltages of less than 15~V. Less than a nanowatt of electrical power is needed to sustain an FSR frequency shift, compatible with the densely packed arrays of microresonators proposed for next generation computer architectures~\cite{atabaki2018integrating}. Our devices are fabricated from silica-on-silicon allowing arbitrary resonance frequency tuning across the full silica transparency window. They could be translated straightforwardly into other material platforms, such as silicon nitride, silicon-on-insulator and CMOS depending on application~\cite{atabaki2018integrating}. To illustrate the capabilities of full FSR tuning, we demonstrate a two microcavity reconfigurable photonic network consisting of one FSR tunable device brought on resonance with a fixed-frequency microtoroidal cavity across a telecommunications optical fibre link. 

We expect that full FSR electrical tuning of high quality optical microcavities will enable a broad range of science and applications, from high-bandwidth optical interconnects and memories in next generation computer architectures~\cite{atabaki2018integrating,kuramochi2014large}, to on-chip tuneable filters, lasers and coherent radar systems~\cite{ellis_photonic_2017,polman2004ultralow,li2013low, ghelfi2014fully, xue2018microcomb, zhuang2007single}, reconfigurable sensor networks~\cite{iqbal2010label}, quantum networks~\cite{konoike2016demand, elshaari2017chip, aoki_observation_2006}, and arrays of nonlinear photonic systems to study collective phenomena such as phase transitions and topological behavior~\cite{heinrich2011collective, douglas2015quantum, gil-santos_light-mediated_2017}.

\bibliography{FSRPaper}

\begin{thebibliography}{48}%
\makeatletter
\providecommand \@ifxundefined [1]{%
 \@ifx{#1\undefined}
}%
\providecommand \@ifnum [1]{%
 \ifnum #1\expandafter \@firstoftwo
 \else \expandafter \@secondoftwo
 \fi
}%
\providecommand \@ifx [1]{%
 \ifx #1\expandafter \@firstoftwo
 \else \expandafter \@secondoftwo
 \fi
}%
\providecommand \natexlab [1]{#1}%
\providecommand \enquote  [1]{``#1''}%
\providecommand \bibnamefont  [1]{#1}%
\providecommand \bibfnamefont [1]{#1}%
\providecommand \citenamefont [1]{#1}%
\providecommand \href@noop [0]{\@secondoftwo}%
\providecommand \href [0]{\begingroup \@sanitize@url \@href}%
\providecommand \@href[1]{\@@startlink{#1}\@@href}%
\providecommand \@@href[1]{\endgroup#1\@@endlink}%
\providecommand \@sanitize@url [0]{\catcode `\\12\catcode `\$12\catcode
  `\&12\catcode `\#12\catcode `\^12\catcode `\_12\catcode `\%12\relax}%
\providecommand \@@startlink[1]{}%
\providecommand \@@endlink[0]{}%
\providecommand \url  [0]{\begingroup\@sanitize@url \@url }%
\providecommand \@url [1]{\endgroup\@href {#1}{\urlprefix }}%
\providecommand \urlprefix  [0]{URL }%
\providecommand \Eprint [0]{\href }%
\providecommand \doibase [0]{http://dx.doi.org/}%
\providecommand \selectlanguage [0]{\@gobble}%
\providecommand \bibinfo  [0]{\@secondoftwo}%
\providecommand \bibfield  [0]{\@secondoftwo}%
\providecommand \translation [1]{[#1]}%
\providecommand \BibitemOpen [0]{}%
\providecommand \bibitemStop [0]{}%
\providecommand \bibitemNoStop [0]{.\EOS\space}%
\providecommand \EOS [0]{\spacefactor3000\relax}%
\providecommand \BibitemShut  [1]{\csname bibitem#1\endcsname}%
\let\auto@bib@innerbib\@empty
\bibitem [{\citenamefont {Atabaki}\ \emph {et~al.}(2018)\citenamefont
  {Atabaki}, \citenamefont {Moazeni}, \citenamefont {Pavanello}, \citenamefont
  {Gevorgyan}, \citenamefont {Notaros}, \citenamefont {Alloatti}, \citenamefont
  {Wade}, \citenamefont {Sun}, \citenamefont {Kruger}, \citenamefont {Meng}
  \emph {et~al.}}]{atabaki2018integrating}%
  \BibitemOpen
  \bibfield  {author} {\bibinfo {author} {\bibfnamefont {A.~H.}\ \bibnamefont
  {Atabaki}}, \bibinfo {author} {\bibfnamefont {S.}~\bibnamefont {Moazeni}},
  \bibinfo {author} {\bibfnamefont {F.}~\bibnamefont {Pavanello}}, \bibinfo
  {author} {\bibfnamefont {H.}~\bibnamefont {Gevorgyan}}, \bibinfo {author}
  {\bibfnamefont {J.}~\bibnamefont {Notaros}}, \bibinfo {author} {\bibfnamefont
  {L.}~\bibnamefont {Alloatti}}, \bibinfo {author} {\bibfnamefont {M.~T.}\
  \bibnamefont {Wade}}, \bibinfo {author} {\bibfnamefont {C.}~\bibnamefont
  {Sun}}, \bibinfo {author} {\bibfnamefont {S.~A.}\ \bibnamefont {Kruger}},
  \bibinfo {author} {\bibfnamefont {H.}~\bibnamefont {Meng}},  \emph {et~al.},\
  }\href@noop {} {\bibfield  {journal} {\bibinfo  {journal} {Nature}\ }\textbf
  {\bibinfo {volume} {556}},\ \bibinfo {pages} {349} (\bibinfo {year}
  {2018})}\BibitemShut {NoStop}%
\bibitem [{\citenamefont {Kuramochi}\ \emph {et~al.}(2014)\citenamefont
  {Kuramochi}, \citenamefont {Nozaki}, \citenamefont {Shinya}, \citenamefont
  {Takeda}, \citenamefont {Sato}, \citenamefont {Matsuo}, \citenamefont
  {Taniyama}, \citenamefont {Sumikura},\ and\ \citenamefont
  {Notomi}}]{kuramochi2014large}%
  \BibitemOpen
  \bibfield  {author} {\bibinfo {author} {\bibfnamefont {E.}~\bibnamefont
  {Kuramochi}}, \bibinfo {author} {\bibfnamefont {K.}~\bibnamefont {Nozaki}},
  \bibinfo {author} {\bibfnamefont {A.}~\bibnamefont {Shinya}}, \bibinfo
  {author} {\bibfnamefont {K.}~\bibnamefont {Takeda}}, \bibinfo {author}
  {\bibfnamefont {T.}~\bibnamefont {Sato}}, \bibinfo {author} {\bibfnamefont
  {S.}~\bibnamefont {Matsuo}}, \bibinfo {author} {\bibfnamefont
  {H.}~\bibnamefont {Taniyama}}, \bibinfo {author} {\bibfnamefont
  {H.}~\bibnamefont {Sumikura}}, \ and\ \bibinfo {author} {\bibfnamefont
  {M.}~\bibnamefont {Notomi}},\ }\href@noop {} {\bibfield  {journal} {\bibinfo
  {journal} {Nature Photonics}\ }\textbf {\bibinfo {volume} {8}},\ \bibinfo
  {pages} {474} (\bibinfo {year} {2014})}\BibitemShut {NoStop}%
\bibitem [{\citenamefont {Konoike}\ \emph {et~al.}(2016)\citenamefont
  {Konoike}, \citenamefont {Nakagawa}, \citenamefont {Nakadai}, \citenamefont
  {Asano}, \citenamefont {Tanaka},\ and\ \citenamefont
  {Noda}}]{konoike2016demand}%
  \BibitemOpen
  \bibfield  {author} {\bibinfo {author} {\bibfnamefont {R.}~\bibnamefont
  {Konoike}}, \bibinfo {author} {\bibfnamefont {H.}~\bibnamefont {Nakagawa}},
  \bibinfo {author} {\bibfnamefont {M.}~\bibnamefont {Nakadai}}, \bibinfo
  {author} {\bibfnamefont {T.}~\bibnamefont {Asano}}, \bibinfo {author}
  {\bibfnamefont {Y.}~\bibnamefont {Tanaka}}, \ and\ \bibinfo {author}
  {\bibfnamefont {S.}~\bibnamefont {Noda}},\ }\href@noop {} {\bibfield
  {journal} {\bibinfo  {journal} {Science advances}\ }\textbf {\bibinfo
  {volume} {2}},\ \bibinfo {pages} {e1501690} (\bibinfo {year}
  {2016})}\BibitemShut {NoStop}%
\bibitem [{\citenamefont {Elshaari}\ \emph {et~al.}(2017)\citenamefont
  {Elshaari}, \citenamefont {Zadeh}, \citenamefont {Fognini}, \citenamefont
  {Reimer}, \citenamefont {Dalacu}, \citenamefont {Poole}, \citenamefont
  {Zwiller},\ and\ \citenamefont {J{\"o}ns}}]{elshaari2017chip}%
  \BibitemOpen
  \bibfield  {author} {\bibinfo {author} {\bibfnamefont {A.~W.}\ \bibnamefont
  {Elshaari}}, \bibinfo {author} {\bibfnamefont {I.~E.}\ \bibnamefont {Zadeh}},
  \bibinfo {author} {\bibfnamefont {A.}~\bibnamefont {Fognini}}, \bibinfo
  {author} {\bibfnamefont {M.~E.}\ \bibnamefont {Reimer}}, \bibinfo {author}
  {\bibfnamefont {D.}~\bibnamefont {Dalacu}}, \bibinfo {author} {\bibfnamefont
  {P.~J.}\ \bibnamefont {Poole}}, \bibinfo {author} {\bibfnamefont
  {V.}~\bibnamefont {Zwiller}}, \ and\ \bibinfo {author} {\bibfnamefont
  {K.~D.}\ \bibnamefont {J{\"o}ns}},\ }\href@noop {} {\bibfield  {journal}
  {\bibinfo  {journal} {Nature communications}\ }\textbf {\bibinfo {volume}
  {8}},\ \bibinfo {pages} {379} (\bibinfo {year} {2017})}\BibitemShut {NoStop}%
\bibitem [{\citenamefont {Aoki}\ \emph {et~al.}(2006)\citenamefont {Aoki},
  \citenamefont {Dayan}, \citenamefont {Wilcut}, \citenamefont {Bowen},
  \citenamefont {Parkins}, \citenamefont {Kippenberg}, \citenamefont {Vahala},\
  and\ \citenamefont {Kimble}}]{aoki_observation_2006}%
  \BibitemOpen
  \bibfield  {author} {\bibinfo {author} {\bibfnamefont {T.}~\bibnamefont
  {Aoki}}, \bibinfo {author} {\bibfnamefont {B.}~\bibnamefont {Dayan}},
  \bibinfo {author} {\bibfnamefont {E.}~\bibnamefont {Wilcut}}, \bibinfo
  {author} {\bibfnamefont {W.~P.}\ \bibnamefont {Bowen}}, \bibinfo {author}
  {\bibfnamefont {A.~S.}\ \bibnamefont {Parkins}}, \bibinfo {author}
  {\bibfnamefont {T.~J.}\ \bibnamefont {Kippenberg}}, \bibinfo {author}
  {\bibfnamefont {K.~J.}\ \bibnamefont {Vahala}}, \ and\ \bibinfo {author}
  {\bibfnamefont {H.~J.}\ \bibnamefont {Kimble}},\ }\href {\doibase
  10.1038/nature05147} {\bibfield  {journal} {\bibinfo  {journal} {Nature}\
  }\textbf {\bibinfo {volume} {443}},\ \bibinfo {pages} {671} (\bibinfo {year}
  {2006})}\BibitemShut {NoStop}%
\bibitem [{\citenamefont {Ghelfi}\ \emph {et~al.}(2014)\citenamefont {Ghelfi},
  \citenamefont {Laghezza}, \citenamefont {Scotti}, \citenamefont {Serafino},
  \citenamefont {Capria}, \citenamefont {Pinna}, \citenamefont {Onori},
  \citenamefont {Porzi}, \citenamefont {Scaffardi}, \citenamefont {Malacarne}
  \emph {et~al.}}]{ghelfi2014fully}%
  \BibitemOpen
  \bibfield  {author} {\bibinfo {author} {\bibfnamefont {P.}~\bibnamefont
  {Ghelfi}}, \bibinfo {author} {\bibfnamefont {F.}~\bibnamefont {Laghezza}},
  \bibinfo {author} {\bibfnamefont {F.}~\bibnamefont {Scotti}}, \bibinfo
  {author} {\bibfnamefont {G.}~\bibnamefont {Serafino}}, \bibinfo {author}
  {\bibfnamefont {A.}~\bibnamefont {Capria}}, \bibinfo {author} {\bibfnamefont
  {S.}~\bibnamefont {Pinna}}, \bibinfo {author} {\bibfnamefont
  {D.}~\bibnamefont {Onori}}, \bibinfo {author} {\bibfnamefont
  {C.}~\bibnamefont {Porzi}}, \bibinfo {author} {\bibfnamefont
  {M.}~\bibnamefont {Scaffardi}}, \bibinfo {author} {\bibfnamefont
  {A.}~\bibnamefont {Malacarne}},  \emph {et~al.},\ }\href@noop {} {\bibfield
  {journal} {\bibinfo  {journal} {Nature}\ }\textbf {\bibinfo {volume} {507}},\
  \bibinfo {pages} {341} (\bibinfo {year} {2014})}\BibitemShut {NoStop}%
\bibitem [{\citenamefont {Xue}\ \emph {et~al.}(2018)\citenamefont {Xue},
  \citenamefont {Xuan}, \citenamefont {Bao}, \citenamefont {Li}, \citenamefont
  {Zheng}, \citenamefont {Zhou}, \citenamefont {Qi},\ and\ \citenamefont
  {Weiner}}]{xue2018microcomb}%
  \BibitemOpen
  \bibfield  {author} {\bibinfo {author} {\bibfnamefont {X.}~\bibnamefont
  {Xue}}, \bibinfo {author} {\bibfnamefont {Y.}~\bibnamefont {Xuan}}, \bibinfo
  {author} {\bibfnamefont {C.}~\bibnamefont {Bao}}, \bibinfo {author}
  {\bibfnamefont {S.}~\bibnamefont {Li}}, \bibinfo {author} {\bibfnamefont
  {X.}~\bibnamefont {Zheng}}, \bibinfo {author} {\bibfnamefont
  {B.}~\bibnamefont {Zhou}}, \bibinfo {author} {\bibfnamefont {M.}~\bibnamefont
  {Qi}}, \ and\ \bibinfo {author} {\bibfnamefont {A.~M.}\ \bibnamefont
  {Weiner}},\ }\href@noop {} {\bibfield  {journal} {\bibinfo  {journal}
  {Journal of Lightwave Technology}\ }\textbf {\bibinfo {volume} {36}},\
  \bibinfo {pages} {2312} (\bibinfo {year} {2018})}\BibitemShut {NoStop}%
\bibitem [{\citenamefont {Zhuang}\ \emph {et~al.}(2007)\citenamefont {Zhuang},
  \citenamefont {Roeloffzen}, \citenamefont {Heideman}, \citenamefont
  {Borreman}, \citenamefont {Meijerink},\ and\ \citenamefont {van
  Etten}}]{zhuang2007single}%
  \BibitemOpen
  \bibfield  {author} {\bibinfo {author} {\bibfnamefont {L.}~\bibnamefont
  {Zhuang}}, \bibinfo {author} {\bibfnamefont {C.}~\bibnamefont {Roeloffzen}},
  \bibinfo {author} {\bibfnamefont {R.}~\bibnamefont {Heideman}}, \bibinfo
  {author} {\bibfnamefont {A.}~\bibnamefont {Borreman}}, \bibinfo {author}
  {\bibfnamefont {A.}~\bibnamefont {Meijerink}}, \ and\ \bibinfo {author}
  {\bibfnamefont {W.}~\bibnamefont {van Etten}},\ }\href@noop {} {\bibfield
  {journal} {\bibinfo  {journal} {IEEE Photonics Technology Letters}\ }\textbf
  {\bibinfo {volume} {19}},\ \bibinfo {pages} {1130} (\bibinfo {year}
  {2007})}\BibitemShut {NoStop}%
\bibitem [{\citenamefont {Klein}\ \emph {et~al.}(2005)\citenamefont {Klein},
  \citenamefont {Geuzebroek}, \citenamefont {Kelderman}, \citenamefont {Sengo},
  \citenamefont {Baker},\ and\ \citenamefont
  {Driessen}}]{klein_reconfigurable_2005}%
  \BibitemOpen
  \bibfield  {author} {\bibinfo {author} {\bibfnamefont {E.~J.}\ \bibnamefont
  {Klein}}, \bibinfo {author} {\bibfnamefont {D.~H.}\ \bibnamefont
  {Geuzebroek}}, \bibinfo {author} {\bibfnamefont {H.}~\bibnamefont
  {Kelderman}}, \bibinfo {author} {\bibfnamefont {G.}~\bibnamefont {Sengo}},
  \bibinfo {author} {\bibfnamefont {N.}~\bibnamefont {Baker}}, \ and\ \bibinfo
  {author} {\bibfnamefont {A.}~\bibnamefont {Driessen}},\ }\href {\doibase
  10.1109/LPT.2005.858131} {\bibfield  {journal} {\bibinfo  {journal} {IEEE
  Photonics Technology Letters}\ }\textbf {\bibinfo {volume} {17}},\ \bibinfo
  {pages} {2358} (\bibinfo {year} {2005})}\BibitemShut {NoStop}%
\bibitem [{\citenamefont {Polman}\ \emph {et~al.}(2004)\citenamefont {Polman},
  \citenamefont {Min}, \citenamefont {Kalkman}, \citenamefont {Kippenberg},\
  and\ \citenamefont {Vahala}}]{polman2004ultralow}%
  \BibitemOpen
  \bibfield  {author} {\bibinfo {author} {\bibfnamefont {A.}~\bibnamefont
  {Polman}}, \bibinfo {author} {\bibfnamefont {B.}~\bibnamefont {Min}},
  \bibinfo {author} {\bibfnamefont {J.}~\bibnamefont {Kalkman}}, \bibinfo
  {author} {\bibfnamefont {T.}~\bibnamefont {Kippenberg}}, \ and\ \bibinfo
  {author} {\bibfnamefont {K.}~\bibnamefont {Vahala}},\ }\href@noop {}
  {\bibfield  {journal} {\bibinfo  {journal} {Applied Physics Letters}\
  }\textbf {\bibinfo {volume} {84}},\ \bibinfo {pages} {1037} (\bibinfo {year}
  {2004})}\BibitemShut {NoStop}%
\bibitem [{\citenamefont {Li}\ \emph {et~al.}(2013)\citenamefont {Li},
  \citenamefont {Xiao}, \citenamefont {Yan}, \citenamefont {Clements},\ and\
  \citenamefont {Gong}}]{li2013low}%
  \BibitemOpen
  \bibfield  {author} {\bibinfo {author} {\bibfnamefont {B.-B.}\ \bibnamefont
  {Li}}, \bibinfo {author} {\bibfnamefont {Y.-F.}\ \bibnamefont {Xiao}},
  \bibinfo {author} {\bibfnamefont {M.-Y.}\ \bibnamefont {Yan}}, \bibinfo
  {author} {\bibfnamefont {W.~R.}\ \bibnamefont {Clements}}, \ and\ \bibinfo
  {author} {\bibfnamefont {Q.}~\bibnamefont {Gong}},\ }\href@noop {} {\bibfield
   {journal} {\bibinfo  {journal} {Optics letters}\ }\textbf {\bibinfo {volume}
  {38}},\ \bibinfo {pages} {1802} (\bibinfo {year} {2013})}\BibitemShut
  {NoStop}%
\bibitem [{\citenamefont {Lu}\ \emph {et~al.}(2011)\citenamefont {Lu},
  \citenamefont {Lee}, \citenamefont {Chen}, \citenamefont {Herchak},
  \citenamefont {Kim}, \citenamefont {Fraser}, \citenamefont {Flagan},\ and\
  \citenamefont {Vahala}}]{lu2011high}%
  \BibitemOpen
  \bibfield  {author} {\bibinfo {author} {\bibfnamefont {T.}~\bibnamefont
  {Lu}}, \bibinfo {author} {\bibfnamefont {H.}~\bibnamefont {Lee}}, \bibinfo
  {author} {\bibfnamefont {T.}~\bibnamefont {Chen}}, \bibinfo {author}
  {\bibfnamefont {S.}~\bibnamefont {Herchak}}, \bibinfo {author} {\bibfnamefont
  {J.-H.}\ \bibnamefont {Kim}}, \bibinfo {author} {\bibfnamefont {S.~E.}\
  \bibnamefont {Fraser}}, \bibinfo {author} {\bibfnamefont {R.~C.}\
  \bibnamefont {Flagan}}, \ and\ \bibinfo {author} {\bibfnamefont
  {K.}~\bibnamefont {Vahala}},\ }\href@noop {} {\bibfield  {journal} {\bibinfo
  {journal} {Proceedings of the National Academy of Sciences}\ }\textbf
  {\bibinfo {volume} {108}},\ \bibinfo {pages} {5976} (\bibinfo {year}
  {2011})}\BibitemShut {NoStop}%
\bibitem [{\citenamefont {Li}\ \emph {et~al.}(2014)\citenamefont {Li},
  \citenamefont {Clements}, \citenamefont {Yu}, \citenamefont {Shi},
  \citenamefont {Gong},\ and\ \citenamefont {Xiao}}]{li2014single}%
  \BibitemOpen
  \bibfield  {author} {\bibinfo {author} {\bibfnamefont {B.-B.}\ \bibnamefont
  {Li}}, \bibinfo {author} {\bibfnamefont {W.~R.}\ \bibnamefont {Clements}},
  \bibinfo {author} {\bibfnamefont {X.-C.}\ \bibnamefont {Yu}}, \bibinfo
  {author} {\bibfnamefont {K.}~\bibnamefont {Shi}}, \bibinfo {author}
  {\bibfnamefont {Q.}~\bibnamefont {Gong}}, \ and\ \bibinfo {author}
  {\bibfnamefont {Y.-F.}\ \bibnamefont {Xiao}},\ }\href@noop {} {\bibfield
  {journal} {\bibinfo  {journal} {Proceedings of the National Academy of
  Sciences}\ }\textbf {\bibinfo {volume} {111}},\ \bibinfo {pages} {14657}
  (\bibinfo {year} {2014})}\BibitemShut {NoStop}%
\bibitem [{\citenamefont {Forstner}\ \emph {et~al.}(2014)\citenamefont
  {Forstner}, \citenamefont {Sheridan}, \citenamefont {Knittel}, \citenamefont
  {Humphreys}, \citenamefont {Brawley}, \citenamefont {Rubinsztein-Dunlop},\
  and\ \citenamefont {Bowen}}]{forstner2014ultrasensitive}%
  \BibitemOpen
  \bibfield  {author} {\bibinfo {author} {\bibfnamefont {S.}~\bibnamefont
  {Forstner}}, \bibinfo {author} {\bibfnamefont {E.}~\bibnamefont {Sheridan}},
  \bibinfo {author} {\bibfnamefont {J.}~\bibnamefont {Knittel}}, \bibinfo
  {author} {\bibfnamefont {C.~L.}\ \bibnamefont {Humphreys}}, \bibinfo {author}
  {\bibfnamefont {G.~A.}\ \bibnamefont {Brawley}}, \bibinfo {author}
  {\bibfnamefont {H.}~\bibnamefont {Rubinsztein-Dunlop}}, \ and\ \bibinfo
  {author} {\bibfnamefont {W.~P.}\ \bibnamefont {Bowen}},\ }\href@noop {}
  {\bibfield  {journal} {\bibinfo  {journal} {Advanced materials}\ }\textbf
  {\bibinfo {volume} {26}},\ \bibinfo {pages} {6348} (\bibinfo {year}
  {2014})}\BibitemShut {NoStop}%
\bibitem [{\citenamefont {Heylman}\ \emph {et~al.}(2017)\citenamefont
  {Heylman}, \citenamefont {Knapper}, \citenamefont {Horak}, \citenamefont
  {Rea}, \citenamefont {Vanga},\ and\ \citenamefont
  {Goldsmith}}]{heylman_optical_2017}%
  \BibitemOpen
  \bibfield  {author} {\bibinfo {author} {\bibfnamefont {K.~D.}\ \bibnamefont
  {Heylman}}, \bibinfo {author} {\bibfnamefont {K.~A.}\ \bibnamefont
  {Knapper}}, \bibinfo {author} {\bibfnamefont {E.~H.}\ \bibnamefont {Horak}},
  \bibinfo {author} {\bibfnamefont {M.~T.}\ \bibnamefont {Rea}}, \bibinfo
  {author} {\bibfnamefont {S.~K.}\ \bibnamefont {Vanga}}, \ and\ \bibinfo
  {author} {\bibfnamefont {R.~H.}\ \bibnamefont {Goldsmith}},\ }\href {\doibase
  10.1002/adma.201700037} {\bibfield  {journal} {\bibinfo  {journal} {Advanced
  Materials}\ ,\ \bibinfo {pages} {1700037}} (\bibinfo {year}
  {2017})}\BibitemShut {NoStop}%
\bibitem [{\citenamefont {Del~Haye}\ \emph {et~al.}(2007)\citenamefont
  {Del~Haye}, \citenamefont {Schliesser}, \citenamefont {Arcizet},
  \citenamefont {Wilken}, \citenamefont {Holzwarth},\ and\ \citenamefont
  {Kippenberg}}]{del2007optical}%
  \BibitemOpen
  \bibfield  {author} {\bibinfo {author} {\bibfnamefont {P.}~\bibnamefont
  {Del~Haye}}, \bibinfo {author} {\bibfnamefont {A.}~\bibnamefont
  {Schliesser}}, \bibinfo {author} {\bibfnamefont {O.}~\bibnamefont {Arcizet}},
  \bibinfo {author} {\bibfnamefont {T.}~\bibnamefont {Wilken}}, \bibinfo
  {author} {\bibfnamefont {R.}~\bibnamefont {Holzwarth}}, \ and\ \bibinfo
  {author} {\bibfnamefont {T.~J.}\ \bibnamefont {Kippenberg}},\ }\href@noop {}
  {\bibfield  {journal} {\bibinfo  {journal} {Nature}\ }\textbf {\bibinfo
  {volume} {450}},\ \bibinfo {pages} {1214} (\bibinfo {year}
  {2007})}\BibitemShut {NoStop}%
\bibitem [{\citenamefont {Shalaev}(2007)}]{shalaev2007optical}%
  \BibitemOpen
  \bibfield  {author} {\bibinfo {author} {\bibfnamefont {V.~M.}\ \bibnamefont
  {Shalaev}},\ }\href@noop {} {\bibfield  {journal} {\bibinfo  {journal}
  {Nature photonics}\ }\textbf {\bibinfo {volume} {1}},\ \bibinfo {pages} {41}
  (\bibinfo {year} {2007})}\BibitemShut {NoStop}%
\bibitem [{\citenamefont {Iqbal}\ \emph {et~al.}(2010)\citenamefont {Iqbal},
  \citenamefont {Gleeson}, \citenamefont {Spaugh}, \citenamefont {Tybor},
  \citenamefont {Gunn}, \citenamefont {Hochberg}, \citenamefont {Baehr-Jones},
  \citenamefont {Bailey},\ and\ \citenamefont {Gunn}}]{iqbal2010label}%
  \BibitemOpen
  \bibfield  {author} {\bibinfo {author} {\bibfnamefont {M.}~\bibnamefont
  {Iqbal}}, \bibinfo {author} {\bibfnamefont {M.~A.}\ \bibnamefont {Gleeson}},
  \bibinfo {author} {\bibfnamefont {B.}~\bibnamefont {Spaugh}}, \bibinfo
  {author} {\bibfnamefont {F.}~\bibnamefont {Tybor}}, \bibinfo {author}
  {\bibfnamefont {W.~G.}\ \bibnamefont {Gunn}}, \bibinfo {author}
  {\bibfnamefont {M.}~\bibnamefont {Hochberg}}, \bibinfo {author}
  {\bibfnamefont {T.}~\bibnamefont {Baehr-Jones}}, \bibinfo {author}
  {\bibfnamefont {R.~C.}\ \bibnamefont {Bailey}}, \ and\ \bibinfo {author}
  {\bibfnamefont {L.~C.}\ \bibnamefont {Gunn}},\ }\href@noop {} {\bibfield
  {journal} {\bibinfo  {journal} {IEEE Journal of Selected Topics in Quantum
  Electronics}\ }\textbf {\bibinfo {volume} {16}},\ \bibinfo {pages} {654}
  (\bibinfo {year} {2010})}\BibitemShut {NoStop}%
\bibitem [{\citenamefont {Heinrich}\ \emph {et~al.}(2011)\citenamefont
  {Heinrich}, \citenamefont {Ludwig}, \citenamefont {Qian}, \citenamefont
  {Kubala},\ and\ \citenamefont {Marquardt}}]{heinrich2011collective}%
  \BibitemOpen
  \bibfield  {author} {\bibinfo {author} {\bibfnamefont {G.}~\bibnamefont
  {Heinrich}}, \bibinfo {author} {\bibfnamefont {M.}~\bibnamefont {Ludwig}},
  \bibinfo {author} {\bibfnamefont {J.}~\bibnamefont {Qian}}, \bibinfo {author}
  {\bibfnamefont {B.}~\bibnamefont {Kubala}}, \ and\ \bibinfo {author}
  {\bibfnamefont {F.}~\bibnamefont {Marquardt}},\ }\href@noop {} {\bibfield
  {journal} {\bibinfo  {journal} {Physical review letters}\ }\textbf {\bibinfo
  {volume} {107}},\ \bibinfo {pages} {043603} (\bibinfo {year}
  {2011})}\BibitemShut {NoStop}%
\bibitem [{\citenamefont {Douglas}\ \emph {et~al.}(2015)\citenamefont
  {Douglas}, \citenamefont {Habibian}, \citenamefont {Hung}, \citenamefont
  {Gorshkov}, \citenamefont {Kimble},\ and\ \citenamefont
  {Chang}}]{douglas2015quantum}%
  \BibitemOpen
  \bibfield  {author} {\bibinfo {author} {\bibfnamefont {J.~S.}\ \bibnamefont
  {Douglas}}, \bibinfo {author} {\bibfnamefont {H.}~\bibnamefont {Habibian}},
  \bibinfo {author} {\bibfnamefont {C.-L.}\ \bibnamefont {Hung}}, \bibinfo
  {author} {\bibfnamefont {A.}~\bibnamefont {Gorshkov}}, \bibinfo {author}
  {\bibfnamefont {H.~J.}\ \bibnamefont {Kimble}}, \ and\ \bibinfo {author}
  {\bibfnamefont {D.~E.}\ \bibnamefont {Chang}},\ }\href@noop {} {\bibfield
  {journal} {\bibinfo  {journal} {Nature Photonics}\ }\textbf {\bibinfo
  {volume} {9}},\ \bibinfo {pages} {326} (\bibinfo {year} {2015})}\BibitemShut
  {NoStop}%
\bibitem [{\citenamefont {Gil-Santos}\ \emph {et~al.}(2017)\citenamefont
  {Gil-Santos}, \citenamefont {Labousse}, \citenamefont {Baker}, \citenamefont
  {Goetschy}, \citenamefont {Hease}, \citenamefont {Gomez}, \citenamefont
  {Lemaître}, \citenamefont {Leo}, \citenamefont {Ciuti},\ and\ \citenamefont
  {Favero}}]{gil-santos_light-mediated_2017}%
  \BibitemOpen
  \bibfield  {author} {\bibinfo {author} {\bibfnamefont {E.}~\bibnamefont
  {Gil-Santos}}, \bibinfo {author} {\bibfnamefont {M.}~\bibnamefont
  {Labousse}}, \bibinfo {author} {\bibfnamefont {C.}~\bibnamefont {Baker}},
  \bibinfo {author} {\bibfnamefont {A.}~\bibnamefont {Goetschy}}, \bibinfo
  {author} {\bibfnamefont {W.}~\bibnamefont {Hease}}, \bibinfo {author}
  {\bibfnamefont {C.}~\bibnamefont {Gomez}}, \bibinfo {author} {\bibfnamefont
  {A.}~\bibnamefont {Lemaître}}, \bibinfo {author} {\bibfnamefont
  {G.}~\bibnamefont {Leo}}, \bibinfo {author} {\bibfnamefont {C.}~\bibnamefont
  {Ciuti}}, \ and\ \bibinfo {author} {\bibfnamefont {I.}~\bibnamefont
  {Favero}},\ }\href {\doibase 10.1103/PhysRevLett.118.063605} {\bibfield
  {journal} {\bibinfo  {journal} {Physical Review Letters}\ }\textbf {\bibinfo
  {volume} {118}},\ \bibinfo {pages} {063605} (\bibinfo {year}
  {2017})}\BibitemShut {NoStop}%
\bibitem [{\citenamefont {Ellis}\ \emph {et~al.}(2017)\citenamefont {Ellis},
  \citenamefont {Kuhlmann}, \citenamefont {Kuehn}, \citenamefont {Spinka},
  \citenamefont {Underwood}, \citenamefont {Gupta}, \citenamefont {Ocola},
  \citenamefont {Liu}, \citenamefont {Wei}, \citenamefont {Stern},
  \citenamefont {Bland-Hawthorn},\ and\ \citenamefont
  {Tuthill}}]{ellis_photonic_2017}%
  \BibitemOpen
  \bibfield  {author} {\bibinfo {author} {\bibfnamefont {S.~C.}\ \bibnamefont
  {Ellis}}, \bibinfo {author} {\bibfnamefont {S.}~\bibnamefont {Kuhlmann}},
  \bibinfo {author} {\bibfnamefont {K.}~\bibnamefont {Kuehn}}, \bibinfo
  {author} {\bibfnamefont {H.}~\bibnamefont {Spinka}}, \bibinfo {author}
  {\bibfnamefont {D.}~\bibnamefont {Underwood}}, \bibinfo {author}
  {\bibfnamefont {R.~R.}\ \bibnamefont {Gupta}}, \bibinfo {author}
  {\bibfnamefont {L.~E.}\ \bibnamefont {Ocola}}, \bibinfo {author}
  {\bibfnamefont {P.}~\bibnamefont {Liu}}, \bibinfo {author} {\bibfnamefont
  {G.}~\bibnamefont {Wei}}, \bibinfo {author} {\bibfnamefont {N.~P.}\
  \bibnamefont {Stern}}, \bibinfo {author} {\bibfnamefont {J.}~\bibnamefont
  {Bland-Hawthorn}}, \ and\ \bibinfo {author} {\bibfnamefont {P.}~\bibnamefont
  {Tuthill}},\ }\href {\doibase 10.1364/OE.25.015868} {\bibfield  {journal}
  {\bibinfo  {journal} {Optics Express}\ }\textbf {\bibinfo {volume} {25}},\
  \bibinfo {pages} {15868} (\bibinfo {year} {2017})}\BibitemShut {NoStop}%
\bibitem [{\citenamefont {Armani}\ \emph {et~al.}(2004)\citenamefont {Armani},
  \citenamefont {Min}, \citenamefont {Martin},\ and\ \citenamefont
  {Vahala}}]{armani_electrical_2004}%
  \BibitemOpen
  \bibfield  {author} {\bibinfo {author} {\bibfnamefont {D.}~\bibnamefont
  {Armani}}, \bibinfo {author} {\bibfnamefont {B.}~\bibnamefont {Min}},
  \bibinfo {author} {\bibfnamefont {A.}~\bibnamefont {Martin}}, \ and\ \bibinfo
  {author} {\bibfnamefont {K.~J.}\ \bibnamefont {Vahala}},\ }\href {\doibase
  10.1063/1.1825069} {\bibfield  {journal} {\bibinfo  {journal} {Applied
  Physics Letters}\ }\textbf {\bibinfo {volume} {85}},\ \bibinfo {pages} {5439}
  (\bibinfo {year} {2004})}\BibitemShut {NoStop}%
\bibitem [{\citenamefont {Jin}\ \emph {et~al.}(2018)\citenamefont {Jin},
  \citenamefont {Polcawich}, \citenamefont {Morton},\ and\ \citenamefont
  {Bowers}}]{jin_piezoelectrically_2018}%
  \BibitemOpen
  \bibfield  {author} {\bibinfo {author} {\bibfnamefont {W.}~\bibnamefont
  {Jin}}, \bibinfo {author} {\bibfnamefont {R.~G.}\ \bibnamefont {Polcawich}},
  \bibinfo {author} {\bibfnamefont {P.~A.}\ \bibnamefont {Morton}}, \ and\
  \bibinfo {author} {\bibfnamefont {J.~E.}\ \bibnamefont {Bowers}},\ }\href
  {\doibase 10.1364/OE.26.003174} {\bibfield  {journal} {\bibinfo  {journal}
  {Opt. Express, OE}\ }\textbf {\bibinfo {volume} {26}},\ \bibinfo {pages}
  {3174} (\bibinfo {year} {2018})}\BibitemShut {NoStop}%
\bibitem [{\citenamefont {Nguyen}\ \emph {et~al.}(2017)\citenamefont {Nguyen},
  \citenamefont {Houwman},\ and\ \citenamefont {Rijnders}}]{nguyen_large_2017}%
  \BibitemOpen
  \bibfield  {author} {\bibinfo {author} {\bibfnamefont {M.~D.}\ \bibnamefont
  {Nguyen}}, \bibinfo {author} {\bibfnamefont {E.~P.}\ \bibnamefont {Houwman}},
  \ and\ \bibinfo {author} {\bibfnamefont {G.}~\bibnamefont {Rijnders}},\
  }\href {\doibase 10.1038/s41598-017-13425-w} {\bibfield  {journal} {\bibinfo
  {journal} {Scientific Reports}\ }\textbf {\bibinfo {volume} {7}},\ \bibinfo
  {pages} {12915} (\bibinfo {year} {2017})}\BibitemShut {NoStop}%
\bibitem [{\citenamefont {Ding}\ \emph {et~al.}(2010)\citenamefont {Ding},
  \citenamefont {Baker}, \citenamefont {Senellart}, \citenamefont {Lemaitre},
  \citenamefont {Ducci}, \citenamefont {Leo},\ and\ \citenamefont
  {Favero}}]{ding_high_2010}%
  \BibitemOpen
  \bibfield  {author} {\bibinfo {author} {\bibfnamefont {L.}~\bibnamefont
  {Ding}}, \bibinfo {author} {\bibfnamefont {C.}~\bibnamefont {Baker}},
  \bibinfo {author} {\bibfnamefont {P.}~\bibnamefont {Senellart}}, \bibinfo
  {author} {\bibfnamefont {A.}~\bibnamefont {Lemaitre}}, \bibinfo {author}
  {\bibfnamefont {S.}~\bibnamefont {Ducci}}, \bibinfo {author} {\bibfnamefont
  {G.}~\bibnamefont {Leo}}, \ and\ \bibinfo {author} {\bibfnamefont
  {I.}~\bibnamefont {Favero}},\ }\href {\doibase
  10.1103/PhysRevLett.105.263903} {\bibfield  {journal} {\bibinfo  {journal}
  {Physical Review Letters}\ }\textbf {\bibinfo {volume} {105}} (\bibinfo
  {year} {2010}),\ 10.1103/PhysRevLett.105.263903}\BibitemShut {NoStop}%
\bibitem [{\citenamefont {Lin}\ \emph {et~al.}(2009)\citenamefont {Lin},
  \citenamefont {Rosenberg}, \citenamefont {Jiang}, \citenamefont {Vahala},\
  and\ \citenamefont {Painter}}]{lin_mechanical_2009}%
  \BibitemOpen
  \bibfield  {author} {\bibinfo {author} {\bibfnamefont {Q.}~\bibnamefont
  {Lin}}, \bibinfo {author} {\bibfnamefont {J.}~\bibnamefont {Rosenberg}},
  \bibinfo {author} {\bibfnamefont {X.}~\bibnamefont {Jiang}}, \bibinfo
  {author} {\bibfnamefont {K.~J.}\ \bibnamefont {Vahala}}, \ and\ \bibinfo
  {author} {\bibfnamefont {O.}~\bibnamefont {Painter}},\ }\href {\doibase
  10.1103/PhysRevLett.103.103601} {\bibfield  {journal} {\bibinfo  {journal}
  {Phys. Rev. Lett.}\ }\textbf {\bibinfo {volume} {103}},\ \bibinfo {pages}
  {103601} (\bibinfo {year} {2009})}\BibitemShut {NoStop}%
\bibitem [{\citenamefont {Baker}\ \emph {et~al.}(2016)\citenamefont {Baker},
  \citenamefont {Bekker}, \citenamefont {McAuslan}, \citenamefont {Sheridan},\
  and\ \citenamefont {Bowen}}]{baker_high_2016}%
  \BibitemOpen
  \bibfield  {author} {\bibinfo {author} {\bibfnamefont {C.~G.}\ \bibnamefont
  {Baker}}, \bibinfo {author} {\bibfnamefont {C.}~\bibnamefont {Bekker}},
  \bibinfo {author} {\bibfnamefont {D.~L.}\ \bibnamefont {McAuslan}}, \bibinfo
  {author} {\bibfnamefont {E.}~\bibnamefont {Sheridan}}, \ and\ \bibinfo
  {author} {\bibfnamefont {W.~P.}\ \bibnamefont {Bowen}},\ }\href {\doibase
  10.1364/OE.24.020400} {\bibfield  {journal} {\bibinfo  {journal} {Optics
  Express}\ }\textbf {\bibinfo {volume} {24}},\ \bibinfo {pages} {20400}
  (\bibinfo {year} {2016})}\BibitemShut {NoStop}%
\bibitem [{\citenamefont {Bekker}\ \emph {et~al.}(2017)\citenamefont {Bekker},
  \citenamefont {Kalra}, \citenamefont {Baker},\ and\ \citenamefont
  {Bowen}}]{bekker_injection_2017}%
  \BibitemOpen
  \bibfield  {author} {\bibinfo {author} {\bibfnamefont {C.}~\bibnamefont
  {Bekker}}, \bibinfo {author} {\bibfnamefont {R.}~\bibnamefont {Kalra}},
  \bibinfo {author} {\bibfnamefont {C.}~\bibnamefont {Baker}}, \ and\ \bibinfo
  {author} {\bibfnamefont {W.~P.}\ \bibnamefont {Bowen}},\ }\href {\doibase
  10.1364/OPTICA.4.001196} {\bibfield  {journal} {\bibinfo  {journal} {OPTICA}\
  }\textbf {\bibinfo {volume} {4}},\ \bibinfo {pages} {1196} (\bibinfo {year}
  {2017})}\BibitemShut {NoStop}%
\bibitem [{\citenamefont {Rosenberg}\ \emph {et~al.}(2009)\citenamefont
  {Rosenberg}, \citenamefont {Lin},\ and\ \citenamefont
  {Painter}}]{rosenberg_static_2009}%
  \BibitemOpen
  \bibfield  {author} {\bibinfo {author} {\bibfnamefont {J.}~\bibnamefont
  {Rosenberg}}, \bibinfo {author} {\bibfnamefont {Q.}~\bibnamefont {Lin}}, \
  and\ \bibinfo {author} {\bibfnamefont {O.}~\bibnamefont {Painter}},\ }\href
  {\doibase 10.1038/nphoton.2009.137} {\bibfield  {journal} {\bibinfo
  {journal} {Nature Photonics}\ }\textbf {\bibinfo {volume} {3}},\ \bibinfo
  {pages} {478} (\bibinfo {year} {2009})}\BibitemShut {NoStop}%
\bibitem [{\citenamefont {Wiederhecker}\ \emph {et~al.}(2009)\citenamefont
  {Wiederhecker}, \citenamefont {Chen}, \citenamefont {Gondarenko},\ and\
  \citenamefont {Lipson}}]{wiederhecker2009controlling}%
  \BibitemOpen
  \bibfield  {author} {\bibinfo {author} {\bibfnamefont {G.~S.}\ \bibnamefont
  {Wiederhecker}}, \bibinfo {author} {\bibfnamefont {L.}~\bibnamefont {Chen}},
  \bibinfo {author} {\bibfnamefont {A.}~\bibnamefont {Gondarenko}}, \ and\
  \bibinfo {author} {\bibfnamefont {M.}~\bibnamefont {Lipson}},\ }\href@noop {}
  {\bibfield  {journal} {\bibinfo  {journal} {nature}\ }\textbf {\bibinfo
  {volume} {462}},\ \bibinfo {pages} {633} (\bibinfo {year}
  {2009})}\BibitemShut {NoStop}%
\bibitem [{\citenamefont {Zhang}\ \emph {et~al.}(2012)\citenamefont {Zhang},
  \citenamefont {Wiederhecker}, \citenamefont {Manipatruni}, \citenamefont
  {Barnard}, \citenamefont {McEuen},\ and\ \citenamefont
  {Lipson}}]{zhang_synchronization_2012}%
  \BibitemOpen
  \bibfield  {author} {\bibinfo {author} {\bibfnamefont {M.}~\bibnamefont
  {Zhang}}, \bibinfo {author} {\bibfnamefont {G.~S.}\ \bibnamefont
  {Wiederhecker}}, \bibinfo {author} {\bibfnamefont {S.}~\bibnamefont
  {Manipatruni}}, \bibinfo {author} {\bibfnamefont {A.}~\bibnamefont
  {Barnard}}, \bibinfo {author} {\bibfnamefont {P.}~\bibnamefont {McEuen}}, \
  and\ \bibinfo {author} {\bibfnamefont {M.}~\bibnamefont {Lipson}},\ }\href
  {\doibase 10.1103/PhysRevLett.109.233906} {\bibfield  {journal} {\bibinfo
  {journal} {Physical Review Letters}\ }\textbf {\bibinfo {volume} {109}},\
  \bibinfo {pages} {233906} (\bibinfo {year} {2012})}\BibitemShut {NoStop}%
\bibitem [{\citenamefont {Lee}\ \emph {et~al.}(2017)\citenamefont {Lee},
  \citenamefont {Zhang}, \citenamefont {Barbosa}, \citenamefont {Miller},
  \citenamefont {Mohanty}, \citenamefont {St-Gelais},\ and\ \citenamefont
  {Lipson}}]{lee_-chip_2017}%
  \BibitemOpen
  \bibfield  {author} {\bibinfo {author} {\bibfnamefont {B.~S.}\ \bibnamefont
  {Lee}}, \bibinfo {author} {\bibfnamefont {M.}~\bibnamefont {Zhang}}, \bibinfo
  {author} {\bibfnamefont {F.~A.~S.}\ \bibnamefont {Barbosa}}, \bibinfo
  {author} {\bibfnamefont {S.~A.}\ \bibnamefont {Miller}}, \bibinfo {author}
  {\bibfnamefont {A.}~\bibnamefont {Mohanty}}, \bibinfo {author} {\bibfnamefont
  {R.}~\bibnamefont {St-Gelais}}, \ and\ \bibinfo {author} {\bibfnamefont
  {M.}~\bibnamefont {Lipson}},\ }\href {\doibase 10.1364/OE.25.012109}
  {\bibfield  {journal} {\bibinfo  {journal} {Opt. Express}\ }\textbf {\bibinfo
  {volume} {25}},\ \bibinfo {pages} {12109} (\bibinfo {year}
  {2017})}\BibitemShut {NoStop}%
\bibitem [{\citenamefont {Sumetsky}\ \emph {et~al.}(2010)\citenamefont
  {Sumetsky}, \citenamefont {Dulashko},\ and\ \citenamefont
  {Windeler}}]{sumetsky_super_2010}%
  \BibitemOpen
  \bibfield  {author} {\bibinfo {author} {\bibfnamefont {M.}~\bibnamefont
  {Sumetsky}}, \bibinfo {author} {\bibfnamefont {Y.}~\bibnamefont {Dulashko}},
  \ and\ \bibinfo {author} {\bibfnamefont {R.~S.}\ \bibnamefont {Windeler}},\
  }\href {https://www.osapublishing.org/abstract.cfm?uri=ol-35-11-1866}
  {\bibfield  {journal} {\bibinfo  {journal} {Optics letters}\ }\textbf
  {\bibinfo {volume} {35}},\ \bibinfo {pages} {1866} (\bibinfo {year}
  {2010})}\BibitemShut {NoStop}%
\bibitem [{\citenamefont {Pöllinger}\ \emph {et~al.}(2009)\citenamefont
  {Pöllinger}, \citenamefont {O’Shea}, \citenamefont {Warken},\ and\
  \citenamefont {Rauschenbeutel}}]{pollinger_ultrahigh-_2009}%
  \BibitemOpen
  \bibfield  {author} {\bibinfo {author} {\bibfnamefont {M.}~\bibnamefont
  {Pöllinger}}, \bibinfo {author} {\bibfnamefont {D.}~\bibnamefont
  {O’Shea}}, \bibinfo {author} {\bibfnamefont {F.}~\bibnamefont {Warken}}, \
  and\ \bibinfo {author} {\bibfnamefont {A.}~\bibnamefont {Rauschenbeutel}},\
  }\href {\doibase 10.1103/PhysRevLett.103.053901} {\bibfield  {journal}
  {\bibinfo  {journal} {Physical Review Letters}\ }\textbf {\bibinfo {volume}
  {103}},\ \bibinfo {pages} {053901} (\bibinfo {year} {2009})}\BibitemShut
  {NoStop}%
\bibitem [{\citenamefont {Chu}\ and\ \citenamefont
  {Hane}(2014)}]{chu_wide_2014}%
  \BibitemOpen
  \bibfield  {author} {\bibinfo {author} {\bibfnamefont {H.~M.}\ \bibnamefont
  {Chu}}\ and\ \bibinfo {author} {\bibfnamefont {K.}~\bibnamefont {Hane}},\
  }\href {\doibase 10.1109/LPT.2014.2326405} {\bibfield  {journal} {\bibinfo
  {journal} {IEEE Photonics Technology Letters}\ }\textbf {\bibinfo {volume}
  {26}},\ \bibinfo {pages} {1411} (\bibinfo {year} {2014})}\BibitemShut
  {NoStop}%
\bibitem [{\citenamefont {Chen}\ \emph {et~al.}(2014)\citenamefont {Chen},
  \citenamefont {Xu}, \citenamefont {Wood},\ and\ \citenamefont
  {Reano}}]{chen_hybrid_2014}%
  \BibitemOpen
  \bibfield  {author} {\bibinfo {author} {\bibfnamefont {L.}~\bibnamefont
  {Chen}}, \bibinfo {author} {\bibfnamefont {Q.}~\bibnamefont {Xu}}, \bibinfo
  {author} {\bibfnamefont {M.~G.}\ \bibnamefont {Wood}}, \ and\ \bibinfo
  {author} {\bibfnamefont {R.~M.}\ \bibnamefont {Reano}},\ }\href {\doibase
  10.1364/OPTICA.1.000112} {\bibfield  {journal} {\bibinfo  {journal} {Optica}\
  }\textbf {\bibinfo {volume} {1}},\ \bibinfo {pages} {112} (\bibinfo {year}
  {2014})}\BibitemShut {NoStop}%
\bibitem [{\citenamefont {Wang}\ \emph {et~al.}(2018)\citenamefont {Wang},
  \citenamefont {Zhang}, \citenamefont {Stern}, \citenamefont {Lipson},\ and\
  \citenamefont {Lončar}}]{wang_nanophotonic_2018}%
  \BibitemOpen
  \bibfield  {author} {\bibinfo {author} {\bibfnamefont {C.}~\bibnamefont
  {Wang}}, \bibinfo {author} {\bibfnamefont {M.}~\bibnamefont {Zhang}},
  \bibinfo {author} {\bibfnamefont {B.}~\bibnamefont {Stern}}, \bibinfo
  {author} {\bibfnamefont {M.}~\bibnamefont {Lipson}}, \ and\ \bibinfo {author}
  {\bibfnamefont {M.}~\bibnamefont {Lončar}},\ }\href {\doibase
  10.1364/OE.26.001547} {\bibfield  {journal} {\bibinfo  {journal} {Optics
  Express}\ }\textbf {\bibinfo {volume} {26}},\ \bibinfo {pages} {1547}
  (\bibinfo {year} {2018})}\BibitemShut {NoStop}%
\bibitem [{\citenamefont {Baker}\ \emph {et~al.}(2014)\citenamefont {Baker},
  \citenamefont {Hease}, \citenamefont {Nguyen}, \citenamefont {Andronico},
  \citenamefont {Ducci}, \citenamefont {Leo},\ and\ \citenamefont
  {Favero}}]{baker_photoelastic_2014}%
  \BibitemOpen
  \bibfield  {author} {\bibinfo {author} {\bibfnamefont {C.}~\bibnamefont
  {Baker}}, \bibinfo {author} {\bibfnamefont {W.}~\bibnamefont {Hease}},
  \bibinfo {author} {\bibfnamefont {D.-T.}\ \bibnamefont {Nguyen}}, \bibinfo
  {author} {\bibfnamefont {A.}~\bibnamefont {Andronico}}, \bibinfo {author}
  {\bibfnamefont {S.}~\bibnamefont {Ducci}}, \bibinfo {author} {\bibfnamefont
  {G.}~\bibnamefont {Leo}}, \ and\ \bibinfo {author} {\bibfnamefont
  {I.}~\bibnamefont {Favero}},\ }\href {\doibase 10.1364/OE.22.014072}
  {\bibfield  {journal} {\bibinfo  {journal} {Optics Express}\ }\textbf
  {\bibinfo {volume} {22}},\ \bibinfo {pages} {14072} (\bibinfo {year}
  {2014})}\BibitemShut {NoStop}%
\bibitem [{\citenamefont {Balram}\ \emph {et~al.}(2014)\citenamefont {Balram},
  \citenamefont {Davan{\c{c}}o}, \citenamefont {Lim}, \citenamefont {Song},\
  and\ \citenamefont {Srinivasan}}]{balram2014moving}%
  \BibitemOpen
  \bibfield  {author} {\bibinfo {author} {\bibfnamefont {K.~C.}\ \bibnamefont
  {Balram}}, \bibinfo {author} {\bibfnamefont {M.}~\bibnamefont
  {Davan{\c{c}}o}}, \bibinfo {author} {\bibfnamefont {J.~Y.}\ \bibnamefont
  {Lim}}, \bibinfo {author} {\bibfnamefont {J.~D.}\ \bibnamefont {Song}}, \
  and\ \bibinfo {author} {\bibfnamefont {K.}~\bibnamefont {Srinivasan}},\
  }\href@noop {} {\bibfield  {journal} {\bibinfo  {journal} {Optica}\ }\textbf
  {\bibinfo {volume} {1}},\ \bibinfo {pages} {414} (\bibinfo {year}
  {2014})}\BibitemShut {NoStop}%
\bibitem [{\citenamefont {Errando-Herranz}\ \emph {et~al.}(2015)\citenamefont
  {Errando-Herranz}, \citenamefont {Niklaus}, \citenamefont {Stemme},\ and\
  \citenamefont {Gylfason}}]{errando-herranz_low-power_2015}%
  \BibitemOpen
  \bibfield  {author} {\bibinfo {author} {\bibfnamefont {C.}~\bibnamefont
  {Errando-Herranz}}, \bibinfo {author} {\bibfnamefont {F.}~\bibnamefont
  {Niklaus}}, \bibinfo {author} {\bibfnamefont {G.}~\bibnamefont {Stemme}}, \
  and\ \bibinfo {author} {\bibfnamefont {K.~B.}\ \bibnamefont {Gylfason}},\
  }\href {\doibase 10.1364/OL.40.003556} {\bibfield  {journal} {\bibinfo
  {journal} {Optics Letters}\ }\textbf {\bibinfo {volume} {40}},\ \bibinfo
  {pages} {3556} (\bibinfo {year} {2015})}\BibitemShut {NoStop}%
\bibitem [{\citenamefont {Jiang}\ \emph {et~al.}(2009)\citenamefont {Jiang},
  \citenamefont {Lin}, \citenamefont {Rosenberg}, \citenamefont {Vahala},\ and\
  \citenamefont {Painter}}]{jiang_high-q_2009}%
  \BibitemOpen
  \bibfield  {author} {\bibinfo {author} {\bibfnamefont {X.}~\bibnamefont
  {Jiang}}, \bibinfo {author} {\bibfnamefont {Q.}~\bibnamefont {Lin}}, \bibinfo
  {author} {\bibfnamefont {J.}~\bibnamefont {Rosenberg}}, \bibinfo {author}
  {\bibfnamefont {K.}~\bibnamefont {Vahala}}, \ and\ \bibinfo {author}
  {\bibfnamefont {O.}~\bibnamefont {Painter}},\ }\href {\doibase
  10.1364/OE.17.020911} {\bibfield  {journal} {\bibinfo  {journal} {Opt.
  Express, OE}\ }\textbf {\bibinfo {volume} {17}},\ \bibinfo {pages} {20911}
  (\bibinfo {year} {2009})}\BibitemShut {NoStop}%
\bibitem [{\citenamefont {Wiederhecker}\ \emph {et~al.}(2011)\citenamefont
  {Wiederhecker}, \citenamefont {Manipatruni}, \citenamefont {Lee},\ and\
  \citenamefont {Lipson}}]{wiederhecker_broadband_2011}%
  \BibitemOpen
  \bibfield  {author} {\bibinfo {author} {\bibfnamefont {G.~S.}\ \bibnamefont
  {Wiederhecker}}, \bibinfo {author} {\bibfnamefont {S.}~\bibnamefont
  {Manipatruni}}, \bibinfo {author} {\bibfnamefont {S.}~\bibnamefont {Lee}}, \
  and\ \bibinfo {author} {\bibfnamefont {M.}~\bibnamefont {Lipson}},\ }\href
  {https://www.osapublishing.org/abstract.cfm?uri=oe-19-3-2782} {\bibfield
  {journal} {\bibinfo  {journal} {Optics Express}\ }\textbf {\bibinfo {volume}
  {19}},\ \bibinfo {pages} {2782} (\bibinfo {year} {2011})}\BibitemShut
  {NoStop}%
\bibitem [{\citenamefont {Aspelmeyer}\ \emph {et~al.}(2014)\citenamefont
  {Aspelmeyer}, \citenamefont {Kippenberg},\ and\ \citenamefont
  {Marquardt}}]{aspelmeyer_cavity_2014}%
  \BibitemOpen
  \bibfield  {author} {\bibinfo {author} {\bibfnamefont {M.}~\bibnamefont
  {Aspelmeyer}}, \bibinfo {author} {\bibfnamefont {T.~J.}\ \bibnamefont
  {Kippenberg}}, \ and\ \bibinfo {author} {\bibfnamefont {F.}~\bibnamefont
  {Marquardt}},\ }\href {\doibase 10.1103/RevModPhys.86.1391} {\bibfield
  {journal} {\bibinfo  {journal} {Reviews of Modern Physics}\ }\textbf
  {\bibinfo {volume} {86}},\ \bibinfo {pages} {1391} (\bibinfo {year}
  {2014})}\BibitemShut {NoStop}%
\bibitem [{Note1()}]{Note1}%
  \BibitemOpen
  \bibinfo {note} {To estimate this, we can compare the typical mechanical
  resonance frequency $\Omega _M$ of the radial breathing mode (tens of MHz)
  for devices of this size to that of the out-of-plane flexural mode (tens of
  kHz). With the spring constant scaling as $k \propto \Omega _M^2$, this gives
  roughly six orders of magnitude larger compliance.}\BibitemShut {Stop}%
\bibitem [{\citenamefont {Thourhout}\ and\ \citenamefont
  {Roels}(2010)}]{thourhout_optomechanical_2010}%
  \BibitemOpen
  \bibfield  {author} {\bibinfo {author} {\bibfnamefont {D.~V.}\ \bibnamefont
  {Thourhout}}\ and\ \bibinfo {author} {\bibfnamefont {J.}~\bibnamefont
  {Roels}},\ }\href {\doibase 10.1038/nphoton.2010.72} {\bibfield  {journal}
  {\bibinfo  {journal} {Nature Photonics}\ }\textbf {\bibinfo {volume} {4}},\
  \bibinfo {pages} {211} (\bibinfo {year} {2010})}\BibitemShut {NoStop}%
\bibitem [{\citenamefont {Baker}\ \emph {et~al.}(2011)\citenamefont {Baker},
  \citenamefont {Belacel}, \citenamefont {Andronico}, \citenamefont
  {Senellart}, \citenamefont {Lemaitre}, \citenamefont {Galopin}, \citenamefont
  {Ducci}, \citenamefont {Leo},\ and\ \citenamefont
  {Favero}}]{baker_critical_2011}%
  \BibitemOpen
  \bibfield  {author} {\bibinfo {author} {\bibfnamefont {C.}~\bibnamefont
  {Baker}}, \bibinfo {author} {\bibfnamefont {C.}~\bibnamefont {Belacel}},
  \bibinfo {author} {\bibfnamefont {A.}~\bibnamefont {Andronico}}, \bibinfo
  {author} {\bibfnamefont {P.}~\bibnamefont {Senellart}}, \bibinfo {author}
  {\bibfnamefont {A.}~\bibnamefont {Lemaitre}}, \bibinfo {author}
  {\bibfnamefont {E.}~\bibnamefont {Galopin}}, \bibinfo {author} {\bibfnamefont
  {S.}~\bibnamefont {Ducci}}, \bibinfo {author} {\bibfnamefont
  {G.}~\bibnamefont {Leo}}, \ and\ \bibinfo {author} {\bibfnamefont
  {I.}~\bibnamefont {Favero}},\ }\href {\doibase 10.1063/1.3651493} {\bibfield
  {journal} {\bibinfo  {journal} {Applied Physics Letters}\ }\textbf {\bibinfo
  {volume} {99}},\ \bibinfo {pages} {151117} (\bibinfo {year}
  {2011})}\BibitemShut {NoStop}%
\bibitem [{Note2()}]{Note2}%
  \BibitemOpen
  \bibinfo {note} {We note here interestingly that SEM measurements cannot be
  relied on for accurate measurements of the disks' geometry and separation,
  because the significant charging brought about by the electron beam creates
  strong electrostatic forces that modify the disk separation and can cause
  collapse of the double-disk structure.}\BibitemShut {Stop}%
\end{thebibliography}%


\begin{thebibliography}{21}%
\makeatletter
\providecommand \@ifxundefined [1]{%
 \@ifx{#1\undefined}
}%
\providecommand \@ifnum [1]{%
 \ifnum #1\expandafter \@firstoftwo
 \else \expandafter \@secondoftwo
 \fi
}%
\providecommand \@ifx [1]{%
 \ifx #1\expandafter \@firstoftwo
 \else \expandafter \@secondoftwo
 \fi
}%
\providecommand \natexlab [1]{#1}%
\providecommand \enquote  [1]{``#1''}%
\providecommand \bibnamefont  [1]{#1}%
\providecommand \bibfnamefont [1]{#1}%
\providecommand \citenamefont [1]{#1}%
\providecommand \href@noop [0]{\@secondoftwo}%
\providecommand \href [0]{\begingroup \@sanitize@url \@href}%
\providecommand \@href[1]{\@@startlink{#1}\@@href}%
\providecommand \@@href[1]{\endgroup#1\@@endlink}%
\providecommand \@sanitize@url [0]{\catcode `\\12\catcode `\$12\catcode
  `\&12\catcode `\#12\catcode `\^12\catcode `\_12\catcode `\%12\relax}%
\providecommand \@@startlink[1]{}%
\providecommand \@@endlink[0]{}%
\providecommand \url  [0]{\begingroup\@sanitize@url \@url }%
\providecommand \@url [1]{\endgroup\@href {#1}{\urlprefix }}%
\providecommand \urlprefix  [0]{URL }%
\providecommand \Eprint [0]{\href }%
\providecommand \doibase [0]{http://dx.doi.org/}%
\providecommand \selectlanguage [0]{\@gobble}%
\providecommand \bibinfo  [0]{\@secondoftwo}%
\providecommand \bibfield  [0]{\@secondoftwo}%
\providecommand \translation [1]{[#1]}%
\providecommand \BibitemOpen [0]{}%
\providecommand \bibitemStop [0]{}%
\providecommand \bibitemNoStop [0]{.\EOS\space}%
\providecommand \EOS [0]{\spacefactor3000\relax}%
\providecommand \BibitemShut  [1]{\csname bibitem#1\endcsname}%
\let\auto@bib@innerbib\@empty
\bibitem [{\citenamefont {Nguyen}\ \emph {et~al.}(2017)\citenamefont {Nguyen},
  \citenamefont {Houwman},\ and\ \citenamefont {Rijnders}}]{nguyen_large_2017}%
  \BibitemOpen
  \bibfield  {author} {\bibinfo {author} {\bibfnamefont {M.~D.}\ \bibnamefont
  {Nguyen}}, \bibinfo {author} {\bibfnamefont {E.~P.}\ \bibnamefont {Houwman}},
  \ and\ \bibinfo {author} {\bibfnamefont {G.}~\bibnamefont {Rijnders}},\
  }\href {\doibase 10.1038/s41598-017-13425-w} {\bibfield  {journal} {\bibinfo
  {journal} {Scientific Reports}\ }\textbf {\bibinfo {volume} {7}},\ \bibinfo
  {pages} {12915} (\bibinfo {year} {2017})}\BibitemShut {NoStop}%
\bibitem [{\citenamefont {Jin}\ \emph {et~al.}(2018)\citenamefont {Jin},
  \citenamefont {Polcawich}, \citenamefont {Morton},\ and\ \citenamefont
  {Bowers}}]{jin_piezoelectrically_2018}%
  \BibitemOpen
  \bibfield  {author} {\bibinfo {author} {\bibfnamefont {W.}~\bibnamefont
  {Jin}}, \bibinfo {author} {\bibfnamefont {R.~G.}\ \bibnamefont {Polcawich}},
  \bibinfo {author} {\bibfnamefont {P.~A.}\ \bibnamefont {Morton}}, \ and\
  \bibinfo {author} {\bibfnamefont {J.~E.}\ \bibnamefont {Bowers}},\ }\href
  {\doibase 10.1364/OE.26.003174} {\bibfield  {journal} {\bibinfo  {journal}
  {Opt. Express, OE}\ }\textbf {\bibinfo {volume} {26}},\ \bibinfo {pages}
  {3174} (\bibinfo {year} {2018})}\BibitemShut {NoStop}%
\bibitem [{Note1()}]{Note1}%
  \BibitemOpen
  \bibinfo {note} {The tuning due to the heat-based increase in cavity size due
  to the material's thermal expansion coefficient is typically an order of
  magnitude smaller than the tuning due to refractive index
  changes.}\BibitemShut {Stop}%
\bibitem [{foo()}]{footnote_1}%
  \BibitemOpen
  \href@noop {} {}\bibinfo {note} {Approximately $1.1\times 10^{-5}$ K$^{-1}$
  and $2.4\times 10^{-5}$ K$^{-1}$ for SiO$_2$ and SiN respectively
  \cite{lee_-chip_2017, baker_optical_2012}.}\BibitemShut {Stop}%
\bibitem [{Note2()}]{Note2}%
  \BibitemOpen
  \bibinfo {note} {Both the electrodes and disks were patterned using electron
  beam lithography (EBL) on a Raith E-Line system with an acceleration voltage
  of 20 kV.}\BibitemShut {Stop}%
\bibitem [{Note3()}]{Note3}%
  \BibitemOpen
  \bibinfo {note} {With CHF$_3$/Ar and SF$_6$/CHF$_3$ chemistry
  respectively.}\BibitemShut {Stop}%
\bibitem [{\citenamefont {Baker}\ \emph {et~al.}(2016)\citenamefont {Baker},
  \citenamefont {Bekker}, \citenamefont {McAuslan}, \citenamefont {Sheridan},\
  and\ \citenamefont {Bowen}}]{baker_high_2016}%
  \BibitemOpen
  \bibfield  {author} {\bibinfo {author} {\bibfnamefont {C.~G.}\ \bibnamefont
  {Baker}}, \bibinfo {author} {\bibfnamefont {C.}~\bibnamefont {Bekker}},
  \bibinfo {author} {\bibfnamefont {D.~L.}\ \bibnamefont {McAuslan}}, \bibinfo
  {author} {\bibfnamefont {E.}~\bibnamefont {Sheridan}}, \ and\ \bibinfo
  {author} {\bibfnamefont {W.~P.}\ \bibnamefont {Bowen}},\ }\href {\doibase
  10.1364/OE.24.020400} {\bibfield  {journal} {\bibinfo  {journal} {Optics
  Express}\ }\textbf {\bibinfo {volume} {24}},\ \bibinfo {pages} {20400}
  (\bibinfo {year} {2016})}\BibitemShut {NoStop}%
\bibitem [{Note4()}]{Note4}%
  \BibitemOpen
  \bibinfo {note} {Or already downwards curved.}\BibitemShut {Stop}%
\bibitem [{\citenamefont {Jiang}\ \emph {et~al.}(2009)\citenamefont {Jiang},
  \citenamefont {Lin}, \citenamefont {Rosenberg}, \citenamefont {Vahala},\ and\
  \citenamefont {Painter}}]{jiang_high-q_2009}%
  \BibitemOpen
  \bibfield  {author} {\bibinfo {author} {\bibfnamefont {X.}~\bibnamefont
  {Jiang}}, \bibinfo {author} {\bibfnamefont {Q.}~\bibnamefont {Lin}}, \bibinfo
  {author} {\bibfnamefont {J.}~\bibnamefont {Rosenberg}}, \bibinfo {author}
  {\bibfnamefont {K.}~\bibnamefont {Vahala}}, \ and\ \bibinfo {author}
  {\bibfnamefont {O.}~\bibnamefont {Painter}},\ }\href {\doibase
  10.1364/OE.17.020911} {\bibfield  {journal} {\bibinfo  {journal} {Opt.
  Express, OE}\ }\textbf {\bibinfo {volume} {17}},\ \bibinfo {pages} {20911}
  (\bibinfo {year} {2009})}\BibitemShut {NoStop}%
\bibitem [{Note5()}]{Note5}%
  \BibitemOpen
  \bibinfo {note} {As well a potential change in the angle of the cantilever at
  the level of its anchoring point.}\BibitemShut {Stop}%
\bibitem [{\citenamefont {Laconte}\ \emph {et~al.}(2006)\citenamefont
  {Laconte}, \citenamefont {Flandre},\ and\ \citenamefont
  {Raskin}}]{laconte_micromachined_2006}%
  \BibitemOpen
  \bibfield  {author} {\bibinfo {author} {\bibfnamefont {J.}~\bibnamefont
  {Laconte}}, \bibinfo {author} {\bibfnamefont {D.}~\bibnamefont {Flandre}}, \
  and\ \bibinfo {author} {\bibfnamefont {J.-P.}\ \bibnamefont {Raskin}},\
  }\href@noop {} {\emph {\bibinfo {title} {Micromachined thin-film sensors for
  {SOI}-{CMOS} co-integration}}}\ (\bibinfo  {publisher} {Springer},\ \bibinfo
  {address} {Dordrecht},\ \bibinfo {year} {2006})\BibitemShut {NoStop}%
\bibitem [{Note6()}]{Note6}%
  \BibitemOpen
  \bibinfo {note} {\protect \textit {i.e.} contact them, at which point they
  are prevented from separating by van der Waals forces. We note that this
  collapse occurs in fabrication during release of the disks through etching of
  the $\alpha $-Si, but not typically during operation.}\BibitemShut {Stop}%
\bibitem [{\citenamefont {Wiederhecker}\ \emph {et~al.}(2011)\citenamefont
  {Wiederhecker}, \citenamefont {Manipatruni}, \citenamefont {Lee},\ and\
  \citenamefont {Lipson}}]{wiederhecker_broadband_2011}%
  \BibitemOpen
  \bibfield  {author} {\bibinfo {author} {\bibfnamefont {G.~S.}\ \bibnamefont
  {Wiederhecker}}, \bibinfo {author} {\bibfnamefont {S.}~\bibnamefont
  {Manipatruni}}, \bibinfo {author} {\bibfnamefont {S.}~\bibnamefont {Lee}}, \
  and\ \bibinfo {author} {\bibfnamefont {M.}~\bibnamefont {Lipson}},\ }\href
  {https://www.osapublishing.org/abstract.cfm?uri=oe-19-3-2782} {\bibfield
  {journal} {\bibinfo  {journal} {Optics Express}\ }\textbf {\bibinfo {volume}
  {19}},\ \bibinfo {pages} {2782} (\bibinfo {year} {2011})}\BibitemShut
  {NoStop}%
\bibitem [{\citenamefont {Iwase}\ \emph {et~al.}(2012)\citenamefont {Iwase},
  \citenamefont {Hui}, \citenamefont {Woolf}, \citenamefont {Rodriguez},
  \citenamefont {Johnson}, \citenamefont {Capasso},\ and\ \citenamefont
  {Lončar}}]{iwase_control_2012}%
  \BibitemOpen
  \bibfield  {author} {\bibinfo {author} {\bibfnamefont {E.}~\bibnamefont
  {Iwase}}, \bibinfo {author} {\bibfnamefont {P.-C.}\ \bibnamefont {Hui}},
  \bibinfo {author} {\bibfnamefont {D.}~\bibnamefont {Woolf}}, \bibinfo
  {author} {\bibfnamefont {A.~W.}\ \bibnamefont {Rodriguez}}, \bibinfo {author}
  {\bibfnamefont {S.~G.}\ \bibnamefont {Johnson}}, \bibinfo {author}
  {\bibfnamefont {F.}~\bibnamefont {Capasso}}, \ and\ \bibinfo {author}
  {\bibfnamefont {M.}~\bibnamefont {Lončar}},\ }\href {\doibase
  10.1088/0960-1317/22/6/065028} {\bibfield  {journal} {\bibinfo  {journal}
  {Journal of Micromechanics and Microengineering}\ }\textbf {\bibinfo {volume}
  {22}},\ \bibinfo {pages} {065028} (\bibinfo {year} {2012})}\BibitemShut
  {NoStop}%
\bibitem [{Note7()}]{Note7}%
  \BibitemOpen
  \bibinfo {note} {The out-of-plane warping for annuli are similar between the
  top and bottom disk, as they are similarly anchored to the central pedestal.
  Hence, though this warping can be on scales of micrometers, the spacing
  between the disks varies much less.}\BibitemShut {Stop}%
\bibitem [{\citenamefont {Wibbeler}\ \emph {et~al.}(1998)\citenamefont
  {Wibbeler}, \citenamefont {Pfeifer},\ and\ \citenamefont
  {Hietschold}}]{wibbeler_parasitic_1998}%
  \BibitemOpen
  \bibfield  {author} {\bibinfo {author} {\bibfnamefont {J.}~\bibnamefont
  {Wibbeler}}, \bibinfo {author} {\bibfnamefont {G.}~\bibnamefont {Pfeifer}}, \
  and\ \bibinfo {author} {\bibfnamefont {M.}~\bibnamefont {Hietschold}},\
  }\href {\doibase 10.1016/S0924-4247(98)00155-1} {\bibfield  {journal}
  {\bibinfo  {journal} {Sensors and Actuators A: Physical}\ }\textbf {\bibinfo
  {volume} {71}},\ \bibinfo {pages} {74} (\bibinfo {year} {1998})}\BibitemShut
  {NoStop}%
\bibitem [{\citenamefont {Bahl}\ \emph {et~al.}(2010)\citenamefont {Bahl},
  \citenamefont {Melamud}, \citenamefont {Kim}, \citenamefont {Chandorkar},
  \citenamefont {Salvia}, \citenamefont {Hopcroft}, \citenamefont {Elata},
  \citenamefont {Hennessy}, \citenamefont {Candler}, \citenamefont {Howe},\
  and\ \citenamefont {Kenny}}]{bahl_model_2010}%
  \BibitemOpen
  \bibfield  {author} {\bibinfo {author} {\bibfnamefont {G.}~\bibnamefont
  {Bahl}}, \bibinfo {author} {\bibfnamefont {R.}~\bibnamefont {Melamud}},
  \bibinfo {author} {\bibfnamefont {B.}~\bibnamefont {Kim}}, \bibinfo {author}
  {\bibfnamefont {S.~A.}\ \bibnamefont {Chandorkar}}, \bibinfo {author}
  {\bibfnamefont {J.~C.}\ \bibnamefont {Salvia}}, \bibinfo {author}
  {\bibfnamefont {M.~A.}\ \bibnamefont {Hopcroft}}, \bibinfo {author}
  {\bibfnamefont {D.}~\bibnamefont {Elata}}, \bibinfo {author} {\bibfnamefont
  {R.~G.}\ \bibnamefont {Hennessy}}, \bibinfo {author} {\bibfnamefont {R.~N.}\
  \bibnamefont {Candler}}, \bibinfo {author} {\bibfnamefont {R.~T.}\
  \bibnamefont {Howe}}, \ and\ \bibinfo {author} {\bibfnamefont {T.~W.}\
  \bibnamefont {Kenny}},\ }\href {\doibase 10.1109/JMEMS.2009.2036274}
  {\bibfield  {journal} {\bibinfo  {journal} {Journal of Microelectromechanical
  Systems}\ }\textbf {\bibinfo {volume} {19}},\ \bibinfo {pages} {162}
  (\bibinfo {year} {2010})}\BibitemShut {NoStop}%
\bibitem [{\citenamefont {Yuan}\ \emph {et~al.}(2004)\citenamefont {Yuan},
  \citenamefont {Cherepko}, \citenamefont {Hwang}, \citenamefont {Goldsmith},
  \citenamefont {Nordqusit},\ and\ \citenamefont {Dyck}}]{yuan_initial_2004}%
  \BibitemOpen
  \bibfield  {author} {\bibinfo {author} {\bibfnamefont {X.}~\bibnamefont
  {Yuan}}, \bibinfo {author} {\bibfnamefont {S.}~\bibnamefont {Cherepko}},
  \bibinfo {author} {\bibfnamefont {J.}~\bibnamefont {Hwang}}, \bibinfo
  {author} {\bibfnamefont {C.~L.}\ \bibnamefont {Goldsmith}}, \bibinfo {author}
  {\bibfnamefont {C.}~\bibnamefont {Nordqusit}}, \ and\ \bibinfo {author}
  {\bibfnamefont {C.}~\bibnamefont {Dyck}},\ }in\ \href {\doibase
  10.1109/MWSYM.2004.1338990} {\emph {\bibinfo {booktitle} {2004 {IEEE}
  {MTT}-{S} {International} {Microwave} {Symposium} {Digest} ({IEEE} {Cat}.
  {No}.04CH37535)}}},\ Vol.~\bibinfo {volume} {3}\ (\bibinfo {year} {2004})\
  pp.\ \bibinfo {pages} {1943--1946}\BibitemShut {NoStop}%
\bibitem [{\citenamefont {Zhou}\ \emph {et~al.}(2016)\citenamefont {Zhou},
  \citenamefont {He}, \citenamefont {He}, \citenamefont {Yu},\ and\
  \citenamefont {Peng}}]{zhou_dielectric_2016}%
  \BibitemOpen
  \bibfield  {author} {\bibinfo {author} {\bibfnamefont {W.}~\bibnamefont
  {Zhou}}, \bibinfo {author} {\bibfnamefont {J.}~\bibnamefont {He}}, \bibinfo
  {author} {\bibfnamefont {X.}~\bibnamefont {He}}, \bibinfo {author}
  {\bibfnamefont {H.}~\bibnamefont {Yu}}, \ and\ \bibinfo {author}
  {\bibfnamefont {B.}~\bibnamefont {Peng}},\ }\href {\doibase
  10.1016/j.microrel.2016.09.004} {\bibfield  {journal} {\bibinfo  {journal}
  {Microelectronics Reliability}\ }\textbf {\bibinfo {volume} {66}},\ \bibinfo
  {pages} {1} (\bibinfo {year} {2016})}\BibitemShut {NoStop}%
\bibitem [{\citenamefont {Lee}\ \emph {et~al.}(2017)\citenamefont {Lee},
  \citenamefont {Zhang}, \citenamefont {Barbosa}, \citenamefont {Miller},
  \citenamefont {Mohanty}, \citenamefont {St-Gelais},\ and\ \citenamefont
  {Lipson}}]{lee_-chip_2017}%
  \BibitemOpen
  \bibfield  {author} {\bibinfo {author} {\bibfnamefont {B.~S.}\ \bibnamefont
  {Lee}}, \bibinfo {author} {\bibfnamefont {M.}~\bibnamefont {Zhang}}, \bibinfo
  {author} {\bibfnamefont {F.~A.~S.}\ \bibnamefont {Barbosa}}, \bibinfo
  {author} {\bibfnamefont {S.~A.}\ \bibnamefont {Miller}}, \bibinfo {author}
  {\bibfnamefont {A.}~\bibnamefont {Mohanty}}, \bibinfo {author} {\bibfnamefont
  {R.}~\bibnamefont {St-Gelais}}, \ and\ \bibinfo {author} {\bibfnamefont
  {M.}~\bibnamefont {Lipson}},\ }\href {\doibase 10.1364/OE.25.012109}
  {\bibfield  {journal} {\bibinfo  {journal} {Opt. Express}\ }\textbf {\bibinfo
  {volume} {25}},\ \bibinfo {pages} {12109} (\bibinfo {year}
  {2017})}\BibitemShut {NoStop}%
\bibitem [{\citenamefont {Baker}\ \emph {et~al.}(2012)\citenamefont {Baker},
  \citenamefont {Stapfner}, \citenamefont {Parrain}, \citenamefont {Ducci},
  \citenamefont {Leo}, \citenamefont {Weig},\ and\ \citenamefont
  {Favero}}]{baker_optical_2012}%
  \BibitemOpen
  \bibfield  {author} {\bibinfo {author} {\bibfnamefont {C.}~\bibnamefont
  {Baker}}, \bibinfo {author} {\bibfnamefont {S.}~\bibnamefont {Stapfner}},
  \bibinfo {author} {\bibfnamefont {D.}~\bibnamefont {Parrain}}, \bibinfo
  {author} {\bibfnamefont {S.}~\bibnamefont {Ducci}}, \bibinfo {author}
  {\bibfnamefont {G.}~\bibnamefont {Leo}}, \bibinfo {author} {\bibfnamefont
  {E.~M.}\ \bibnamefont {Weig}}, \ and\ \bibinfo {author} {\bibfnamefont
  {I.}~\bibnamefont {Favero}},\ }\href {\doibase 10.1364/OE.20.029076}
  {\bibfield  {journal} {\bibinfo  {journal} {Optics Express}\ }\textbf
  {\bibinfo {volume} {20}},\ \bibinfo {pages} {29076} (\bibinfo {year}
  {2012})}\BibitemShut {NoStop}%
\end{thebibliography}%


%

\vspace{3mm}
\textbf{Acknowledgments}
This research was primarily funded by the Australian Research Council and Lockheed Martin Corporation through the Australian Research Council Linkage Grant LP140100595. Support was also provided by a Lockheed Martin Corporation seed grant and the Australian Research Council Centre of Research Excellence for Engineered Quantum Systems (CE110001013). W.P.B., C.G.B, R.K. and B.L. acknowledge fellowships from the Australian Research Council (FT140100650) and the University of Queensland (UQFEL1833877 \&  UQFEL1719237 \& UQFEL14001447), respectively. This work was performed in part at the Queensland node of the Australian National Fabrication Facility, a company established under the National Collaborative Research Infrastructure Strategy to provide nano and microfabrication facilities for Australia's researchers. The authors acknowledge the facilities, and the scientific and technical assistance, of the Australian Microscopy \& Microanalysis Research Facility at the Centre for Microscopy and Microanalysis, The University of Queensland. The authors acknowledge Mariusz Martyniuk and Dhirendra Tripathi at the University of Western Australia for the growth of the wafers, as well as Miaoqiang Liu from the Wang group and Daniel Szombati from the Fedorov group at the University of Queensland, for access to equipment and aid in annealing the wafers and depositing the metal layer, respectively.

\end{document}


\title{Free spectral range electrical tuning of a high quality on-chip microcavity: methods and supplementary information }

\author{Christiaan Bekker}
\thanks{CGB and CB contributed equally \\ Corresponding author: c.baker3@uq.edu.au}
\author{Christopher G. Baker}
\thanks{CGB and CB contributed equally \\ Corresponding author: c.baker3@uq.edu.au}
\author{Rachpon Kalra}
\affiliation{Centre for Engineered Quantum Systems, School of Mathematics and Physics, The University of Queensland, Australia}
\author{Han-Hao Cheng}
\affiliation{Centre for Engineered Quantum Systems, School of Mathematics and Physics, The University of Queensland, Australia}
\affiliation{Centre for Microscopy and Microanalysis, The University of Queensland, Australia}
\author{Bei-Bei Li}
\affiliation{Centre for Engineered Quantum Systems, School of Mathematics and Physics, The University of Queensland, Australia}
\author{Varun Prakash}
\affiliation{Centre for Engineered Quantum Systems, School of Mathematics and Physics, The University of Queensland, Australia}
\author{Warwick P. Bowen}
\affiliation{Centre for Engineered Quantum Systems, School of Mathematics and Physics, The University of Queensland, Australia}

\renewcommand\thesubsection{S\arabic{subsection}}
\setcounter{page}{1}
\renewcommand*{\thepage}{S\arabic{page}}
\setcounter{equation}{0}
\renewcommand{\theequation}{S\arabic{equation}}
\setcounter{figure}{0}
\renewcommand{\thefigure}{S\arabic{figure}}
\renewcommand*{\citenumfont}[1]{S#1}
\renewcommand*{\bibnumfmt}[1]{[S#1]}

\maketitle

\textbf{These supplements contain a discussion of the scaling of thermal and strain-based tuning methods with resonator dimensions (section  \ref{Supp:FSRScaling}),
fabrication details (section \ref{Supp:Fabrication}), modeling of the capacitive actuation scheme (section \ref{Supp:capacitiveactuation}), a discussion of the  mechanical nonlinearities arising from stress in the disk layers (section \ref{Supp:stressandmechnonlinearities}), details on the experimental setup (section \ref{Supp:Experiment}), as well as discussion of improved designs (section \ref{Supp:improveddesign}).}

\subsection{Scaling of tuning methods with device dimensions} 
\label{Supp:FSRScaling}

\begin{figure}[b]
\centering
 \includegraphics[width=0.9\linewidth]{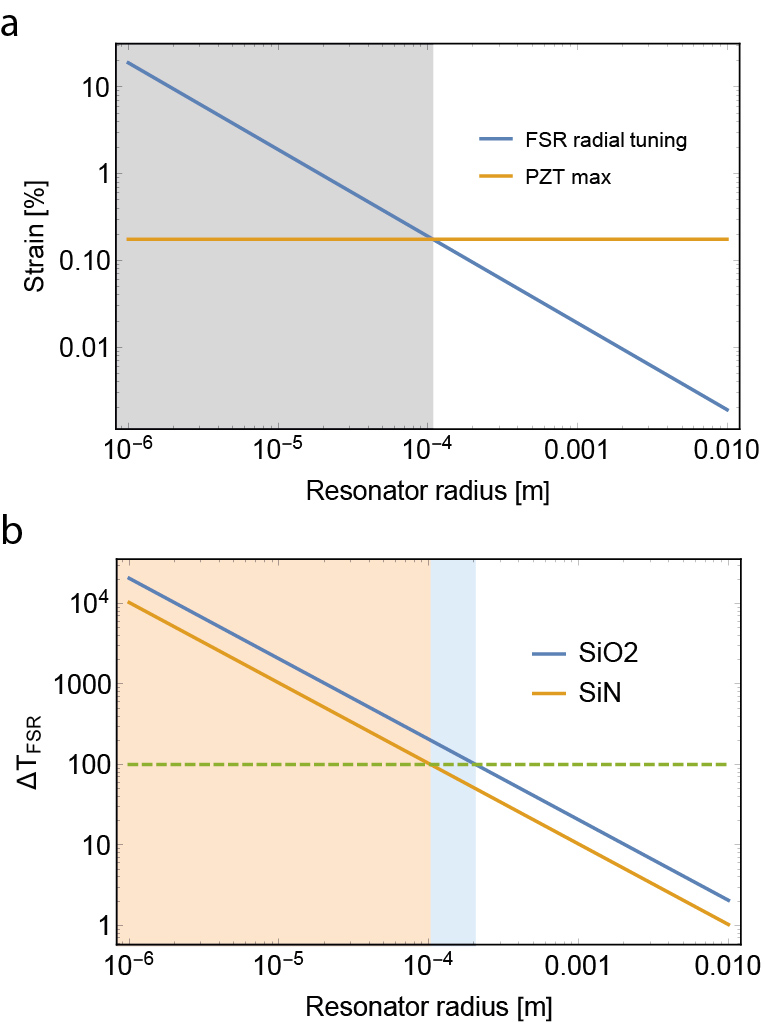}
 \caption{\label{FSRComparison} Scaling of radial-strain (a) and heat-based (b) FSR tuning requirements with device dimensions. }
\end{figure}
 
Figure \ref{FSRComparison} shows how the radial-strain and temperature change $\Delta T_{\text{FSR}}$ requirements for tuning an optical resonator by an FSR become increasingly difficult to achieve with reduced resonator dimensions. Figure \ref{FSRComparison} (a) displays the radial strain $\frac{\Delta R_\text{FSR}}{R}=\frac{\lambda}{2 \pi n_\text{neff} R}$ required for FSR-tuning of a circular resonator, versus resonator dimensions (blue line).  The same absolute change in circumference (one optical wavelength), and hence the same absolute change in radius is required to tune differently-sized devices by a free-spectral-range, leading to a $\frac{1}{R}$ dependence for the required strain. Hence smaller devices require increasingly more strain to tune by a FSR, and at a device radius below $\sim$100 $\mu$m (grey area of Fig \ref{FSRComparison}(a)), this requirement exceeds the maximum strain of common piezoelectric materials such as PZT (0.175\% \cite{nguyen_large_2017} - orange line). Even neglecting the important issues of efficiently communicating the strain between the piezoelectric and resonator layers, this ultimately limits the feasibility of approaches whereby the resonator is strain-tuned by a deposited piezoelectric layer \cite{jin_piezoelectrically_2018} without also engineering a large strain-dependent effective refractive index, as demonstrated in our paper.

Heat-based tuning mainly relies on changing the refractive index of the optical cavity through the thermooptic effect\footnote{The tuning due to the heat-based increase in cavity size due to the material's thermal expansion coefficient is typically an order of magnitude smaller than the tuning due to refractive index changes. }. As the required change in refractive index $\Delta n_\text{eff,\,FSR}=\frac{n_\text{eff}}{n}\left(\frac{\mathrm{d} n}{\mathrm{d} T}\right) \Delta T_\text{FSR}=\frac{\lambda}{2\pi R}$ scales as $\frac{1}{R}$, the temperature change required to tune by an FSR has the same scaling. Further, as the thermo-optic coefficient $\frac{\mathrm{d} n}{\mathrm{d} T}$ is often quite small \cite{footnote_1}
, very large temperatures are required for tuning devices with small dimensions. 
Figure \ref{FSRComparison}(b) plots the temperature increase $\Delta T_\text{FSR}$ required for thermal tuning of a resonator by an FSR, versus resonator dimensions, for resonators made from SiO$_2$ (blue) and SiN (orange). 
Blue and orange regions indicate device dimensions below which FSR thermal tuning is no longer possible,  assuming a maximum temperature increase of the device $\Delta T_\text{max}=100$ K, for SiO$_2$ and SiN respectively. Given this temperature cap, only silicon nitride devices with a radius larger than $\sim$100 $\mu$m can be FSR-tunable, or silica devices with radii exceeding $\sim$200 $\mu$m. 

Though the choice of $\Delta T_\text{max}$ made here is somewhat arbitrary, the absolute limits of thermal tuning are not far away in terms of device size. Indeed, temperature increases in excess of 1000 degrees Celsius ---close to the melting point of glass and above the sublimation temperature of most metal coatings--- would be required at device dimensions of tens of microns. 

Both plots illustrate how radial-strain and heat-based tuning techniques reach their limits for device dimensions typically on the order of hundreds of micrometers. While the capacitive effective-index tuning method presented in this work also becomes increasingly difficult as device dimensions are decreased and FSR is increased, the method is not yet at its limit and can in fact be implemented with devices much smaller than the other tuning methods, allowing for more compact integration, see discussion in section \ref{Supp:improveddesign}.

\subsection{Fabrication}
\label{Supp:Fabrication}

\begin{figure} 
\centering
 \includegraphics[width=\linewidth]{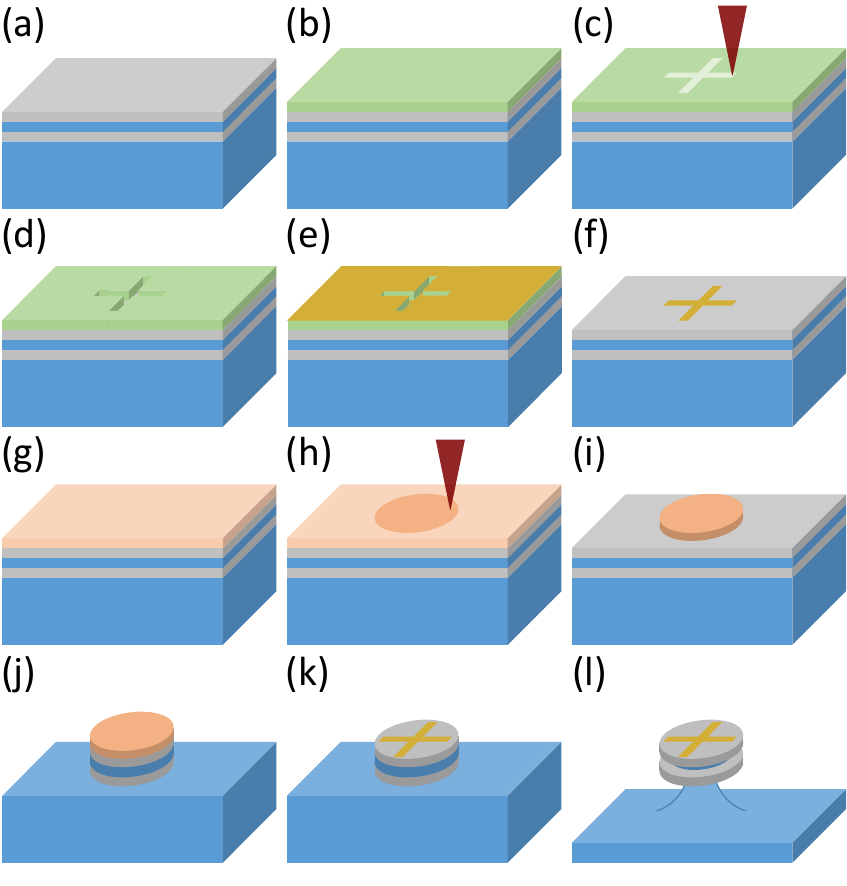}
 \caption{\label{Fig:Fabrication} Fabrication process for double disk electro-optomechanical devices used in this work. Steps are outlined in the text. SiO$_2$ layers are shown in gray; $\alpha$-Si and Si in blue.}
\end{figure}

The wafers used here were made by growing a stack of SiO$_2$/$\alpha$-Si/SiO$_2$ with thickness 350/300/350 nm through ICP-CVD on a Si wafer. The bulk stress of the silica layers was measured to be 120 MPa (compressive). The fabrication for these devices was conducted in two stages: the electrodes were patterned first, and the disks were patterned and etched out afterwards\footnote{Both the electrodes and disks were patterned using electron beam lithography (EBL) on a Raith E-Line system with an acceleration voltage of 20 kV.}. The outline of the fabrication process can be seen in figure \ref{Fig:Fabrication}. 

The electrodes (Fig. \ref{Fig:Fabrication} \textbf{b-f}) were patterned with a bilayer of positive photoresist (PMMA), to allow at once high resolution from the upper layer (PMMA 950K (A4)) during patterning (\textbf{c}) and a large undercut from the lower layer (PMMA 495K (A2)) for lift-off . After evaporation of 10 nm of titanium and 50 nm of gold (\textbf{e}), the photoresist and excess gold was lifted off using N-methylpyrrolidone (NMP) (\textbf{f}). The time required to lift off the metal was greatly reduced by heating the NMP to 70 degrees Celsius. 

In constrast to the electrodes, the disks (Fig. \ref{Fig:Fabrication} \textbf{g-l}) were patterned using a negative photoresist (maN-2410) (\textbf{g}). The stacks were etched out anisotropically using a single run of reactive ion etching, with alternate recipes loaded sequentially to selectively etch  through SiO$_2$ and $\alpha$-Si\footnote{with CHF$_3$/Ar and SF$_6$/CHF$_3$ chemistry respectively.} (\textbf{j}). The $\alpha$-Si sacrificial layer and substrate were simultaneously undercut using a XeF$_2$ isotropic dry etch (30,000:1 selectivity for Si:SiO$_2$, step \textbf{l}).

\subsection{Capacitive actuation}
\label{Supp:capacitiveactuation}

\begin{figure}
\centering
 \includegraphics[width=\linewidth]{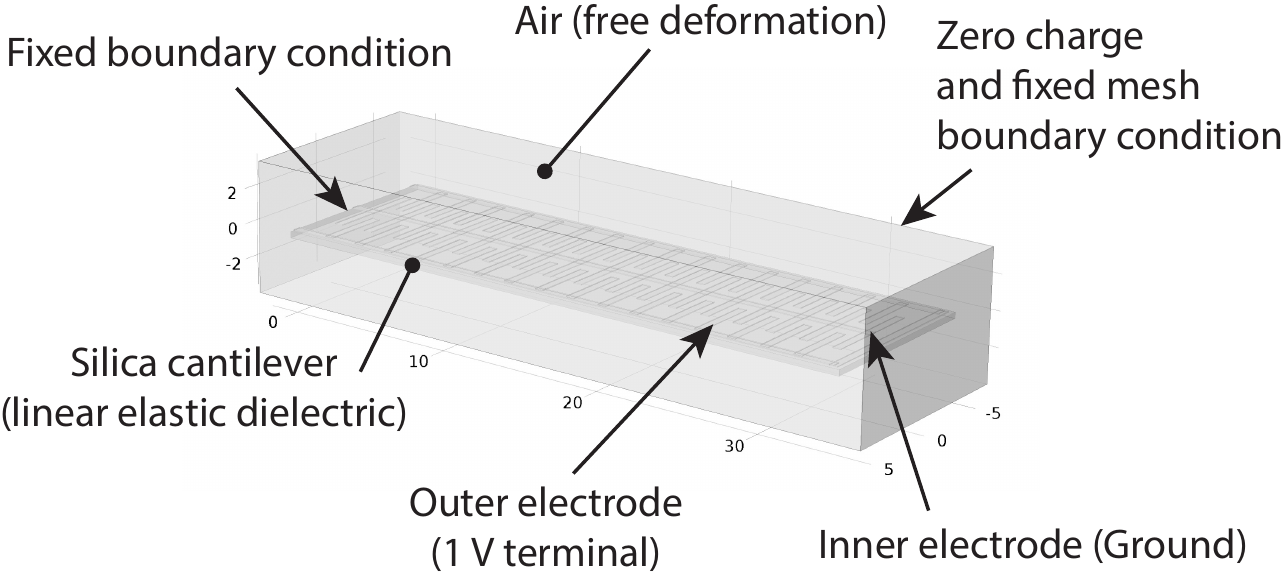}
 \caption{\label{Fig:3Delectromechanics} 3D electromechanical simulation of a support-spoke deflection. The box around the cantilever is an air domain inside which the electrical field lines are calculated, with free mesh deformation (\textit{i.e.} through which the cantilever can deflect without resistance). Calculated $\alpha_\text{mech}=1.1$ nm/V$^2$.}
\end{figure}

Figure \ref{Fig:3Delectromechanics} shows the simulation domain and settings used for the 3D electromechanical simulation employed to calculate the response of the support anchors to a voltage applied to the interdigitated electrodes. Simulations are done with the finite element modelling software Comsol Multiphysics. The support beam is modelled as a cantilever with a fixed (left) and free (right) boundary condition. Interdigitated electrodes on the cantilever surface (electrode width: 500 nm; capacitive gap between electrodes: 500 nm)
are connected to a ground (0 V) and a (1 V) terminal. The new equilibrium position of the spoke is calculated in the presence of this voltage bias. In response to the bias, the system will seek to to maximize its capacitance. This can occur either through a physical narrowing of the electrode gap, or through pulling in more dielectric material between the electrodes \cite{baker_high_2016}. The first effect can be achieved through an upwards curling of the cantilever, effectively bringing the electrodes closer to each other, while the second effect  is achieved through a downwards deflection of the beam, such that the electrodes "envelop" the beam and field lines are more confined to the dielectric \cite{baker_high_2016}.
Our simulations show that the later mechanism is dominant for a straight\footnote{or already downwards curved.} spoke, which will deflect downwards, with a mechanical tunability $\alpha_\text{mech}$ on the order of 1 nm of physical deflection per V$^2$. However, as can be intuited, an initially upwards curved beam will keep curling up upon the application of bias voltage such that the physical narrowing of the electrode gap mechanism dominates. Thus the direction of the mechanical deflection will depend on the initial curvature of a given device. This can be seen by comparing Figure 2 of the main text to Figure \ref{Fig:TuningSupp}, where the device has a slightly different geometry and release process. While the wavelength of the former is tuned down upon application of a voltage bias, corresponding to the top disk deflecting upwards and increasing disk separation, the latter displays a wavelength increase with increasing voltage, as the top disk is deflected downwards. Both deflection directions ultimately lead to similar optical tunability $\alpha_\text{opt}$.

\subsection{Bulk and gradient stresses and mechanical buckling}
\label{Supp:stressandmechnonlinearities}

\subsubsection*{Stress in silica layers} \label{Supp:Warping}

\begin{figure}
\centering
 \includegraphics[width=\linewidth]{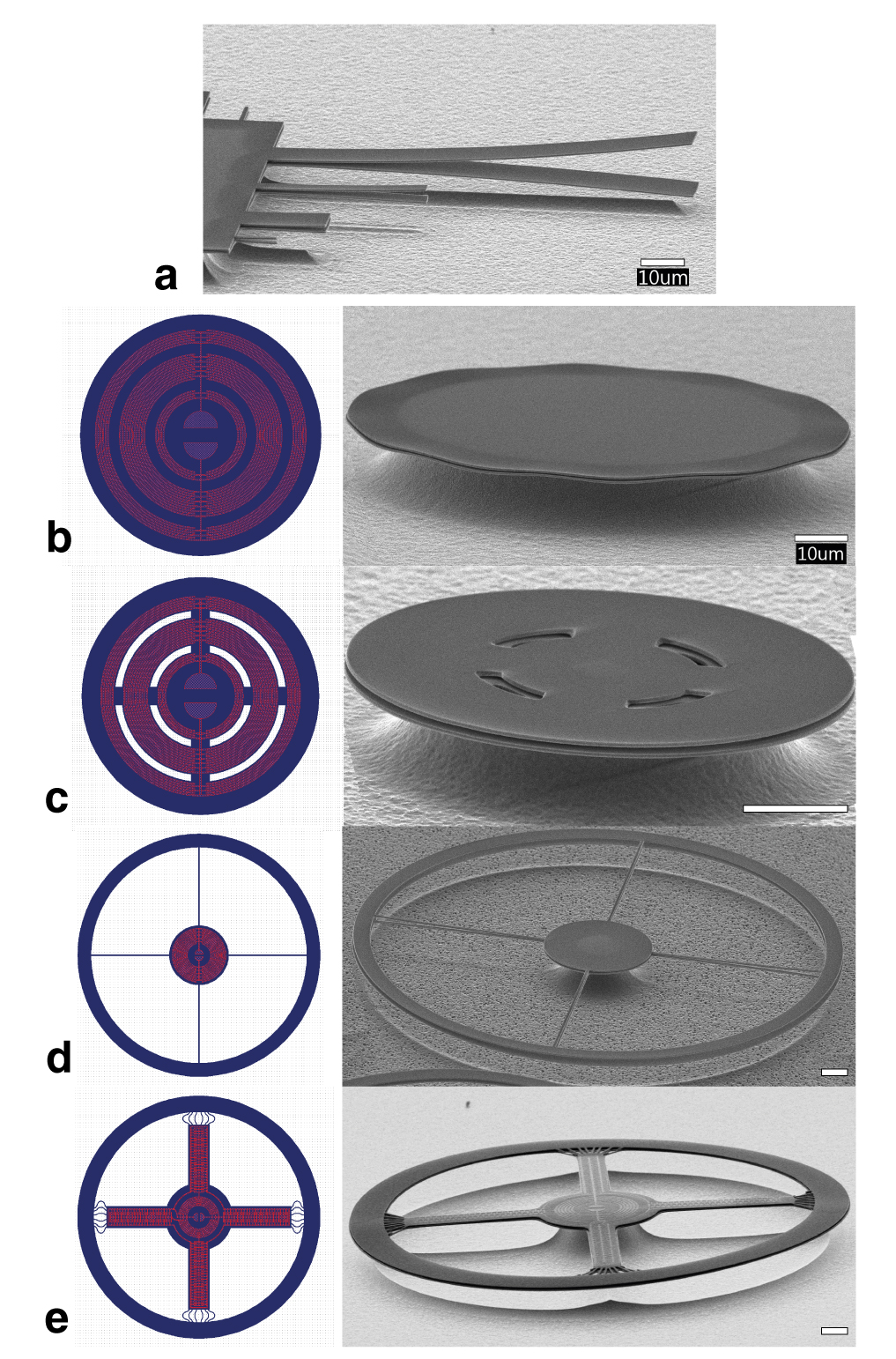}
 \caption{\label{Fig:Warping} (a) Cantilevers fabricated in the silica layers display deformation due to an internal stress gradient in the material. This is the main limitation in the development and design of the double disk cavities in this paper. Sections (b)-(e) show design files and SEM images of the main design steps from the first full-disk design to stress-released stacked annuli. The scale bars correspond to 10 um in each SEM image.}
\end{figure}

Our devices are based on the double disk geometry developed by Jiang,\textit{ et al.} \cite{jiang_high-q_2009}. Aside from the bulk stress in the grown layers of the device, a key design constraint was the internal stress gradient (\emph{i.e.} varying stress in the vertical direction) created in the silica layers during deposition. This caused warping in the structures once released, which could not be completely eliminated, see e.g. the cantilevers in Fig.\ref{Fig:Warping}(a).
Upon underetching of the sacrificial layer, both bulk-stress and vertical stress gradients will relax through a deformation of the device. The bulk (\textit{i.e.} average) compressive strain will, in the case of a cantilever mostly relax through in-plane expansion of the device \footnote{as well a potential change in the angle of the cantilever at the level of its anchoring point.}.  The vertical stress gradient in the layers will on the other hand relax through an upwards or downwards curling of the beam \cite{laconte_micromachined_2006}, as visible in Fig \ref{Fig:Warping}a, which reveals for instance an increasing compressive stress from top to bottom for the top silica layer, responsible for the upwards curling upon release. (Through a fitting of the released cantilever curvature, we extract a stress variation within the layer on the order of 100 MPa \cite{laconte_micromachined_2006}).

We note here that the following stress-mitigation discussion would not be necessary for devices fabricated from lower stress layers. These stress effects caused our initial full-disk designs (Fig. \ref{Fig:Warping}(b)) to fail, as under-etching the sacrificial silicon layer sufficiently to allow electrical actuation induced sufficient warping to collapse the disks\footnote{\textit{i.e.} contact them, at which point they are prevented from separating by van der Waals forces. We note that this collapse occurs in fabrication during release of the disks through etching of the $\alpha$-Si, but not typically during operation.}. Furthermore, the etch rate of the sacrificial layer gets increasingly slow as the etch progresses, as it gets increasingly difficult for reactants and reaction products to make their way through the high-aspect ratio slot between top and bottom disks. In practice, undercutting of the $\alpha$-Si sacrificial layer much beyond 10 microns is difficult to achieve.

Placing slots in the disk to allow for greater underetching and stress release (Fig. \ref{Fig:Warping}(c)) eliminated warping in smaller devices and provided a greater actuation area for the electrodes, however warping was still too strong in devices with diameters larger than 50 $\mu$m to have high device yields. Larger devices are beneficial for this actuation scheme, as the optomechanical coupling strength only depends on disk separation, and a larger device allows both for better mechanical compliance and a larger electrode area. 

The next step in design was to suspend the outer section of the disk on thin spokes connected to a central pedestal (Fig \ref{Fig:Warping}(d)), similar to work done in the Lipson group \citep{wiederhecker_broadband_2011}. This design allows for high mechanical compliance and a minimisation of warping in the device, however as the electrodes are limited to a small central pedestal (whose size is limited so as to avoid warping which can transfer to the outer annuli), the tuning ability of the device is still not optimal. 

Our final design (Fig. \ref{Fig:Warping}(e)) inserts wide spokes between the central pedestal and the outer annuli, providing ample room to actuate the gap spacing through application of  a capacitive force on the spokes. In order to not warp the outer disk, stress-release features inspired by the work of Iwase, \textit{et al.} \cite{iwase_control_2012} are introduced between the annuli and spokes. These features allow the device to relax in the plane of the disk, while providing sufficient support to keep the outer rings suspended.

\subsubsection*{Mechanical buckling} \label{Supp:Buckling}

\begin{figure}
\centering
 \includegraphics[width=\linewidth]{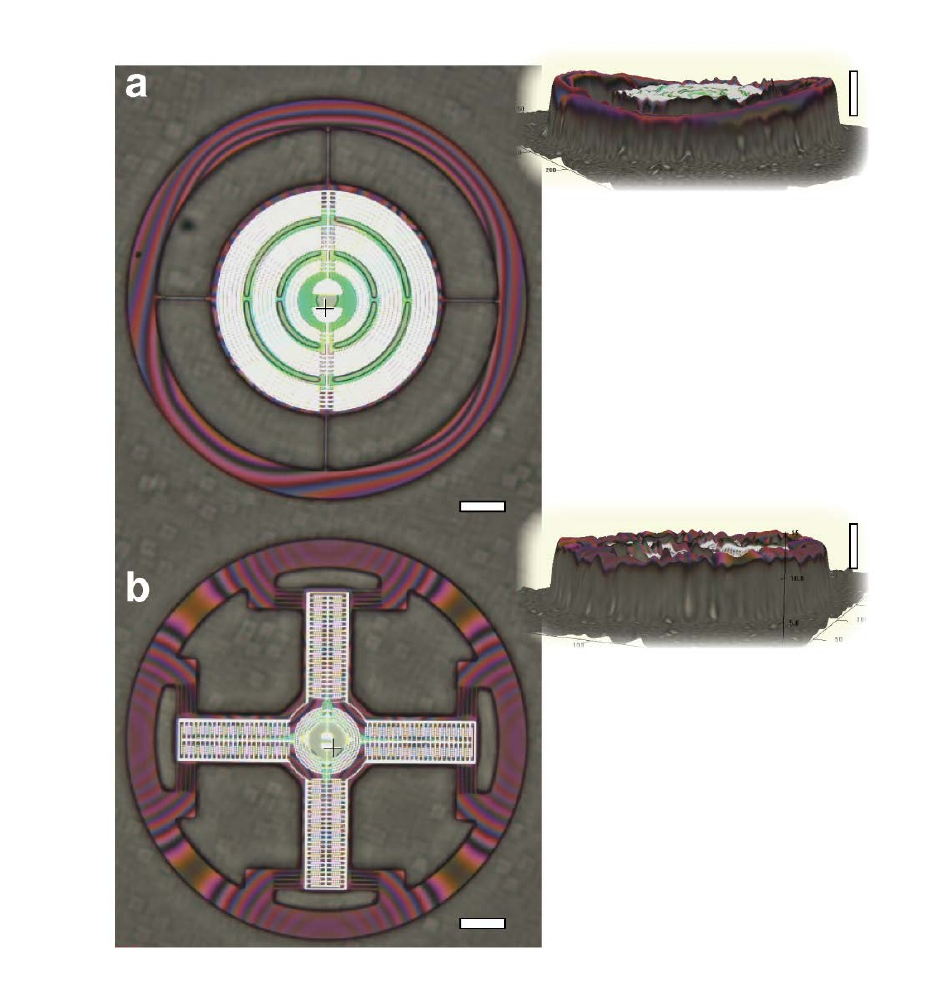}
 \caption{\label{Fig:Buckling} Optical image top-view of a double-disk with thin spokes (a) and double-disk with wide spokes and stress-release features \cite{iwase_control_2012} (b). Visible fringes on the outer annuli are due to optical interference coming from changes in gap spacing. Scale bars are 20 $\mu$m. \textbf{Insets:} 3D optical profiles captured using an optical profiling tool (Zeta 300), showing out-of-plane warping in the devices after release. Vertical scale bars are 5 $\mu$m. Note the significantly reduced out-of-plane displacement in (b) vs (a). }
\end{figure}

Whereas warping in disks caused 'ripples' to appear around the circumference of the device, as visible in Fig. \ref{Fig:Warping}(b), annular designs tended to adopt a saddle-like shape when released, as shown in figure \ref{Fig:Buckling}(a). The stress-release design (Fig. \ref{Fig:Buckling}b) reduced this warping, as the spokes are at once more rigid and allow for radial expansion, minimising the out-of-plane deformation due to stress release\footnote{The out-of-plane warping for annuli are similar between the top and bottom disk, as they are similarly anchored to the central pedestal. Hence, though this warping can be on scales of micrometers, the spacing between the disks varies much less.}. 

We note that, as can be seen in figure \ref{Fig:Buckling}, the integrity of the double-disks could be inspected optically. The presence of interference fringes in the outer annulus indicate changes in the disk separation, and therefore imply that the disks are still physically separated. This was particularly useful once it was realised that imaging the double disks, particularly annular designs, under an SEM at high magnification caused charge build-up in the silica, which could consequently lead to collapse and device failure.

\begin{figure}
\centering
 \includegraphics[width=\linewidth]{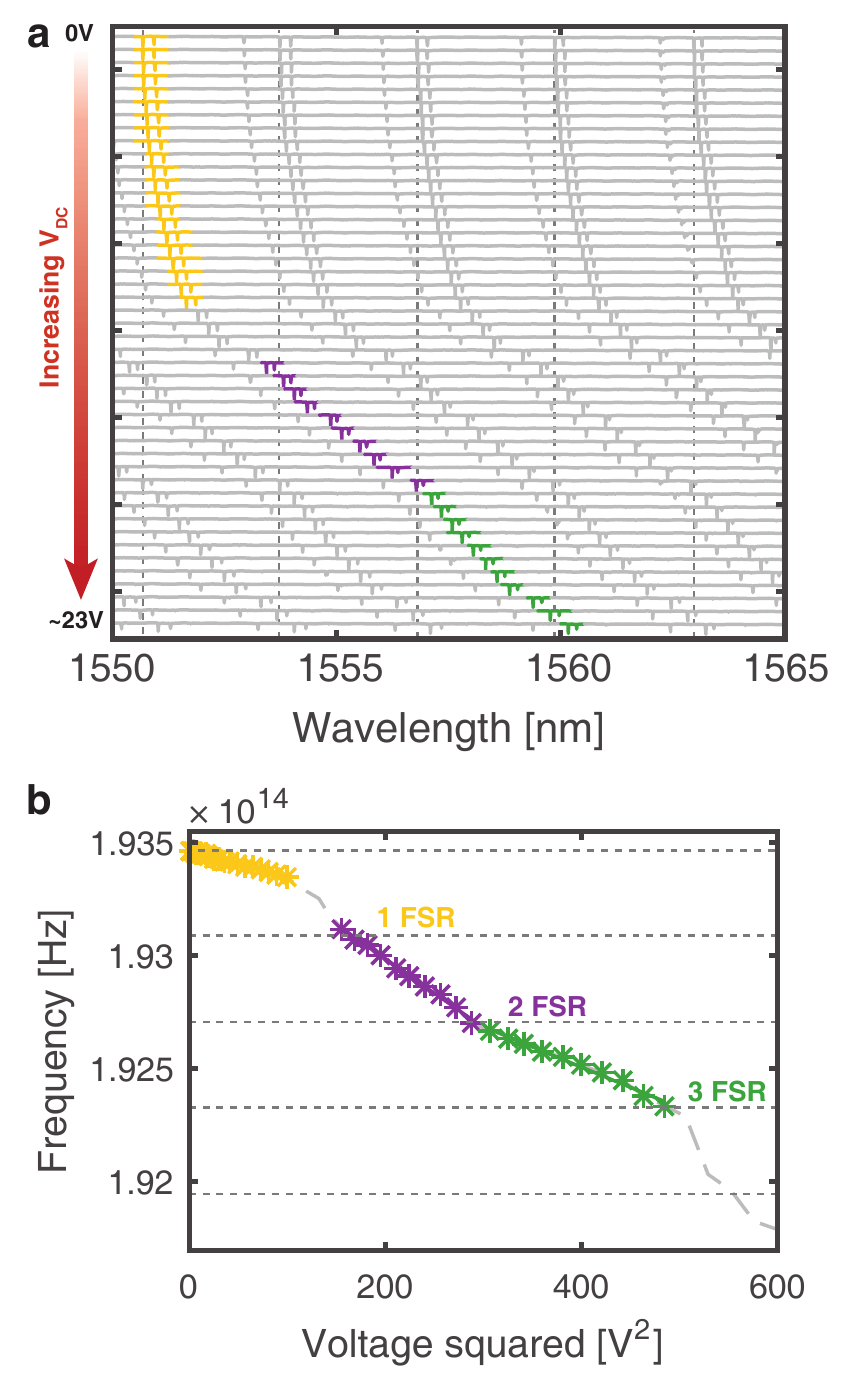}
 \caption{\label{Fig:TuningSupp} (a) Optical tuning of over 3 FSRs, in a different device to the one in shown in Figure 2 of the main text. (b) Clear transitions between regions of different tunability can be seen, with tunabilities of -1.2 (yellow), -3.0 (purple) and -1.8 (green) GHz/V$^2$ respectively.}
\end{figure}

The stress-release features could not completely eliminate the warping however, which is the probable cause of the tunability transitions encountered when tuning the optical resonance (see Fig. 2 of the main text and \ref{Fig:TuningSupp}). Since the spokes may have different curvatures, or even be initially curved in different directions, an applied voltage causes different displacements for each arm, transferring to uneven tuning over the circumference of the device. At some point, the displacement from one arm can affect another sufficiently to buckle it into a different shape, changing the tunability of the device as a whole.
 This mechanism predicts similar transition regions to what was observed in our experiments.  For instance, figure \ref{Fig:TuningSupp} shows the data for optical tuning of a different device to the one measured in the main text, with clear transitions between regions of different tunability (see Fig. \ref{Fig:TuningSupp}(b)). In this device, tuning in excess of 3 FSRs was observed, which could be required for FSR tuning at a larger wavelengths, where the FSR of the device becomes larger. Because these transitions are a feature of the mechanical compliance of the device as it is displaced, transitions are present at the same applied voltages for a given device, and do not pose any problem during operation.
 
Variation in the tunability of different devices was observed due both to the aforementioned mechanical effects and variation in fabrication conditions and geometry. However, tuning on the order $\sim$GHz/V$^2$ was reliably observed in most devices, with the maximal tunability observed being 8.8 GHz/V$^2$. 

\begin{figure} [t!]
\centering
 \includegraphics[width=\linewidth]{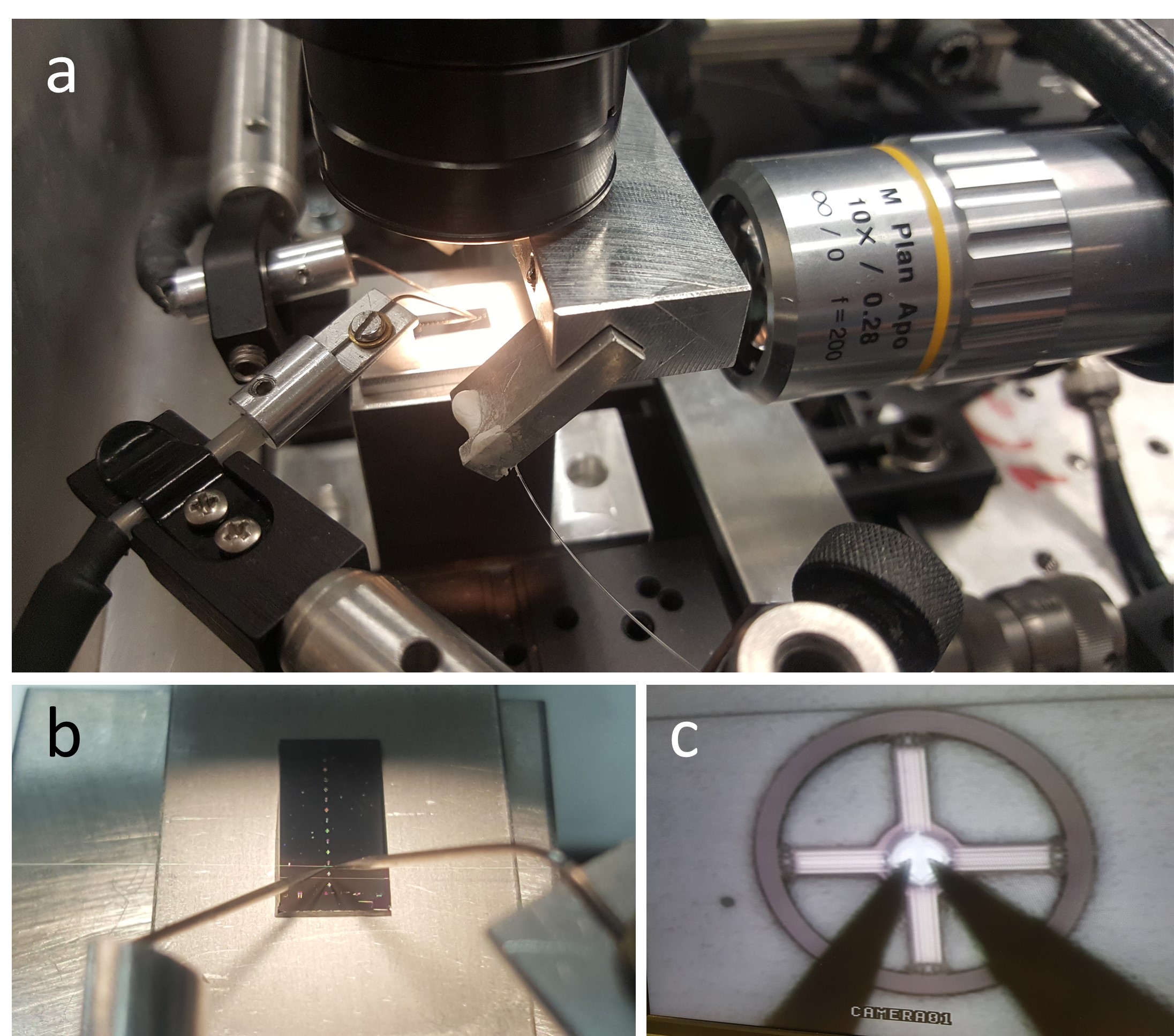}
 \caption{\label{ExSetup_pics} Photographs of the experimental setup used to characterise our devices. (a) Side-view photograph of the experimental setup, showing the 2-axis imaging system. (b) Photograph showing a silicon chip containing 14 devices, as well as two ultra-sharp tungsten probe-tips used for the application of a voltage bias (foreground) and the fibre taper used for evanescent optical readout of the resonators. (c) Optical microscope top-view of a double-disk resonator, with fibre taper (top) and probe tips contacting the device (foreground).}
\end{figure}

\subsubsection*{Charging hysteresis}

We observed some degree of hysteresis with regards to the tuning properties of the device, particularly upon the application of large voltage biases over an extended period of time. We rule out mechanisms coming from device geometry such as buckling transitions, through the observation of similar behaviour on simple cantilever structures with integrated electrodes. Hence it is postulated that this hysteresis is due to charging in the silica disks, a well-known phenomenon in the MEMS community \cite{wibbeler_parasitic_1998, bahl_model_2010, yuan_initial_2004, zhou_dielectric_2016}.  As the tuning method is material agnostic, we note that this problem can be eliminated by using an alternative material for the double disks, for example a semiconductor such as silicon or GaAs.

 \begin{figure} [t!]
\centering
 \includegraphics[width=\linewidth]{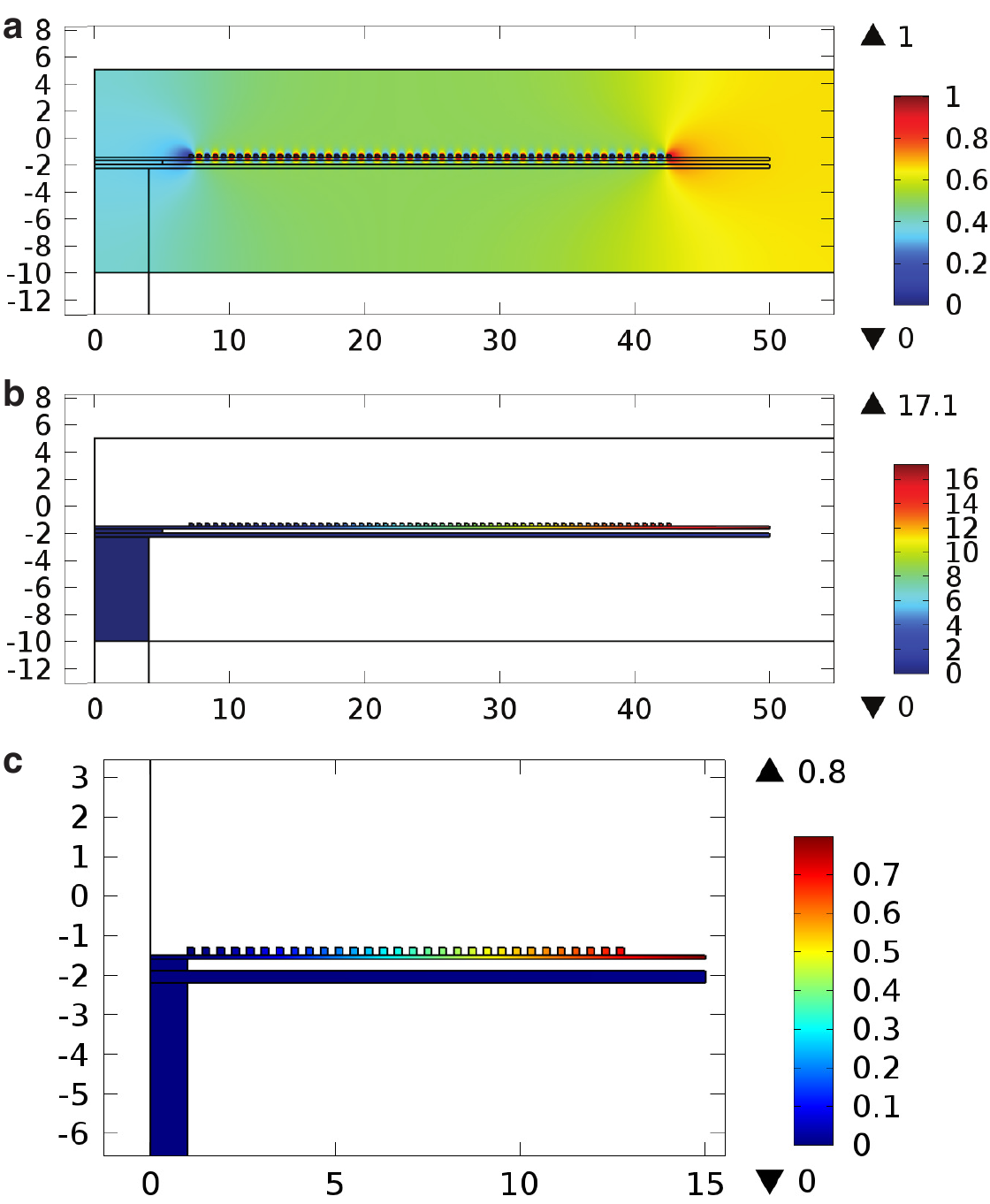}
 \caption{\label{FigImproveddesign} 2D axisymmetric electromechanics simulation of devices with improved tunability. Top:  Calculated electric potential, with one electrode connected to ground and the other to a 1 V source (color code: electric potential (V)). Middle: calculated deflection for the 1 V bias shown above (color code: physical displacement (nm)), yielding $\alpha_\text{mech}=17$ nm/V$^2$. Device radius: 50 um; Electrode width: 300 nm; Electrode gap: 300 nm; N=30 pairs of concentric electrodes. Top and bottom disk thickness respectively 200 nm and 300 nm. Bottom: Deflection for a 1 V bias in a smaller device of radius 15 um (color code :physical displacement), yielding $\alpha_\text{mech}=0.8$ nm/V$^2$. Device radius/surface area are reduced respectively by a factor of 6/36 compared to the devices shown in the main text. Electrode width: 200 nm; Electrode gap: 200 nm; N=15 pairs of concentric electrodes. Top and bottom disk thickness respectively 100 nm and 300 nm.}
\end{figure}

\subsection{Experimental setup} \label{Supp:Experiment}

Figure \ref{ExSetup_pics} shows some pictures of the experimental setup used to probe the devices. The ultrasharp tungsten probe tips (radius of curvature 1 $\mu$m) are each mounted on a 3-axis NanoMax stage. The tapered fibre is also mounted on a 3-axis NanoMax stage, while the device is stationary. Microscopes and piezoelectric position control are used to provide precise electrical contact with the electrode pads on the double-disk.  

\subsection{Improved design}
\label{Supp:improveddesign}

Here we discuss the potential of the tuning approach for further improvement in optical tunability. The main limitation to further gains in $\alpha_\text{opt}$ with existing devices stems from the poorly controlled stress in the silica layers. A reduced stress in the material would enable both a better-controlled and smaller air gap (and hence a higher $G_\text{OM}\sim25$ GHz/nm, see Fig. 1 of the main text), as well as allow for a denser electrode pattern to be deposited on a larger portion of the structure. Combined, these would enable two orders of magnitude gains in optical tunability ---with $\alpha_{\mathrm{opt}}$ in excess of 400 GHz/V$^2$--- in devices half the current size (as in Fig. \ref{FigImproveddesign}(a) and (b)). These could be tuned by an FSR with only 1.3 V. 

Alternatively, these modifications would also enable a device reduced in size by a factor of six (with a radius of 15 um - Fig. \ref{FigImproveddesign}(c)) to still be FSR tunable using an applied voltage of 13 V, with a tunability of $\alpha_{\mathrm{opt}} \simeq$ 14 GHz/V$^2$.

In the absence of spokes needed for stress-release, the geometry of the device becomes rotationally invariant, and the calculation of $\alpha_\text{mech}$ reduces to a 2D-axisymmetric problem, as shown in Figure \ref{FigImproveddesign}.  

Beyond the removal of spokes, for these predictions we make two simple modifications to the current design.
First, narrower electrodes and electrode gaps allow for a more intense electric field as well as a denser electrode pattern, both leading to greater mechanical tunability. 
Second, top and bottom disk can be made asymmetric (\textit{i.e.} of different thicknesses). Indeed, only the top disk is actuated by the electrodes and needs to be mechanically compliant. The use of a thinner top disk leads to an only marginally smaller $G_\text{OM}$, but dramatically improves the mechanical tunability $\alpha_{\mathrm{mech}}$, resulting in a much larger overall $\alpha_{\mathrm{opt}}$ (see Eq. 3 of the main text).

\begin{figure} [t!]
\centering
\includegraphics[width=\linewidth]{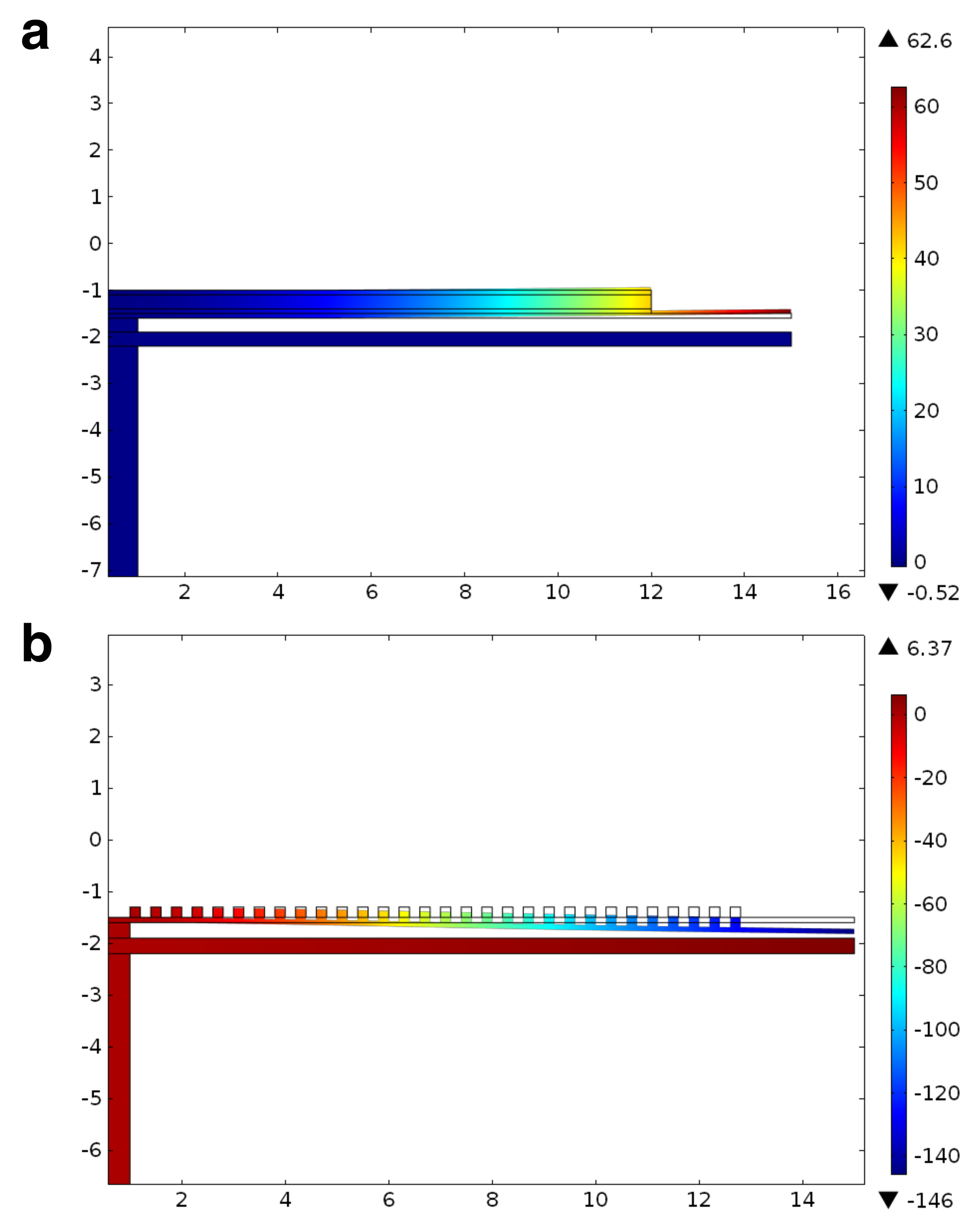}
\caption{\label{Fig:PZTvsElec} 2D axisymmetric FEM simulations comparing piezoelectric (a) and capacitive (b) actuation schemes. Both simulations are for devices with radius 15 um and top/bottom disk thicknesses of 100/300 nm  respectively. (color code: physical displacement (nm)) (a) Piezoelectric scheme similar to Jin, \textit{et al.}\citep{jin_piezoelectrically_2018}, where a piezoelectric element is deposited on top of the device. The out-of-plane displacement of 63 nm results from application of the maximal material strain (2$\times 10^{-3}$). (b) Simulation with the optimised scheme presented in Fig. \ref{FigImproveddesign}(c). As described in the section on improved device design, FSR tuning is possible with this method for an applied voltage of 13~V, as shown in this panel. The resulting out-of-plane displacement of 146 nm is more than twice the maximum achievable with the piezoelectric actuation scheme.}
\end{figure}

\subsubsection*{Comparison with Piezoelectric Tuning}

Here we compare the efficiency of capacitive vs. piezoelectric actuation for double disk resonators. We model piezoelectric actuation for a 15 $\mu$m radius double-disk with a design similar to the work of Jin, \textit{et al.}\citep{jin_piezoelectrically_2018} (see Fig. \ref{Fig:PZTvsElec}(a), to be compared with the model for Fig. \ref{FigImproveddesign}(c)), where a layer of PZT of optimal thickness for actuation (300 nm) is placed between two 100 nm thick planar platinum electrodes (required to bias the PZT) positioned on top of the double-disk. 

Upon expansion/contraction of the PZT layer, the top disk is strained radially, as well as bent out-of-plane. It is this second motion which is responsible for tuning the optical cavity. The optimum in PZT thickness can be understood as follows: An increase in the thickness of the PZT results in more rigidity for the entire structure, and reduces the contribution of the strain in the piezoelectric layer to out-of-plane deflection (and hence $\alpha_\text{opt}$); while reducing the thickness of the PZT will reduce the effectiveness of actuation due to the weaker force applied to the structure. 
The maximum piezoelectric deflection was calculated by giving the PZT layer an initial strain equal to the maximum strain of the material ($2\times10^{-3}$) and allowing it to relax in order to find the stationary displacement of the top disk. This is equivalent to providing the PZT layer with an initial stress sufficient to achieve a final strain with the same magnitude, but opposite sign, to that applied in this model.  

We found that our method of electrical actuation through a set of interdigitated capacitors (shown in Fig. \ref{Fig:PZTvsElec}(b) for the same device dimensions), besides being simpler to fabricate and implement as it only requires a single deposition step, is also more effective at tuning a microcavity at the scale of tens of microns. 
Indeed, in this example with a device radius of 15 $\mu$m, tuning by a full FSR requires the disk separation to be modified by $\sim$140 nm. This displacement is achieved with our capacitive actuation scheme through the application of a 13~V bias (see Fig. \ref{Fig:PZTvsElec}(b)), while it exceeds the maximum out-of-plane displacement of 63 nm achievable with the PZT actuation scheme in our model (Fig. \ref{Fig:PZTvsElec}(a)). 

\bibliography{FSRPapersupplements}